\documentclass[12pt]{article}
\usepackage{graphicx}
\usepackage{epsfig}
\usepackage{latexsym}
\usepackage{latexsym}
\usepackage{amssymb}
\usepackage{amsmath}
\newcommand{\labell}[1]{\label{#1}}
\newcommand{\reef}[1]{(\ref{#1})}

\DeclareSymbolFont{AMSb}{U}{msb}{m}{n}
\DeclareMathSymbol{\IN}{\mathbin}{AMSb}{"4E}
\DeclareMathSymbol{\IZ}{\mathbin}{AMSb}{"5A}
\DeclareMathSymbol{\IR}{\mathbin}{AMSb}{"52}
\DeclareMathSymbol{\Q}{\mathbin}{AMSb}{"51}
\DeclareMathSymbol{\II}{\mathbin}{AMSb}{"49}
\DeclareMathSymbol{\IC}{\mathbin}{AMSb}{"43}
\DeclareMathSymbol{\IP}{\mathbin}{AMSb}{"50}
\DeclareMathSymbol{\IH}{\mathbin}{AMSb}{"48}
\DeclareMathSymbol\IA{\mathalpha}{AMSb}{"41}
\DeclareMathSymbol\IS{\mathalpha}{AMSb}{"53}

\def\Q{{\cal Q}}

\begin{document}

\begin{flushright}
\end{flushright}
\begin{center} {\Large \bf D--Branes  and Fluxes}

\bigskip

{\Large\bf in}

\bigskip

{\Large\bf Supersymmetric Quantum Mechanics}

\end{center}

\bigskip \bigskip \bigskip

\centerline{\bf James E. Carlisle${}^\flat$, Clifford V. Johnson${}^{\natural,}$\footnote{Also, Visiting Professor at the Centre for Particle Theory${}^\flat.$}, Jeffrey S. Pennington${}^\sharp$}

\bigskip
\bigskip

  \centerline{\it ${}^{\flat}$Centre for Particle Theory}
\centerline{\it Department of Mathematical
    Sciences}
\centerline{\it University of Durham}
\centerline{\it Durham DH1 3LE, England, U.K.}

\bigskip
\bigskip

  \centerline{\it ${}^{\natural,\sharp}$Department of Physics and Astronomy }
\centerline{\it University of
Southern California}
\centerline{\it Los Angeles, CA 90089-0484, U.S.A.}

\bigskip
\bigskip

\centerline{\small \tt j.e.carlisle@durham.ac.uk, johnson1@usc.edu, jspennin@usc.edu}

\bigskip
\bigskip


\begin{abstract}

  Type 0A string theory in the $(2,4k)$ superconformal minimal model
  backgrounds, with background ZZ D--branes or R--R fluxes can be
  formulated non--perturbatively. The branes and fluxes have a
  description as threshold bound states in an associated
  one--dimensional quantum mechanics which has a supersymmetric
  structure, familiar from studies of the generalized KdV system. The
  relevant bound state wavefunctions in this problem have unusual
  asymptotics (they are not normalizable in general, and break
  supersymmetry) which are consistent with the underlying description
  in terms of open and closed string sectors.  The overall
  organization of the physics is very pleasing: The physics of the
  closed strings in the background of branes or fluxes is captured by
  the generalized KdV system and non--perturbative string equations
  obtained by reduction of that system (the hierarchy of equations
  found by Dalley, Johnson, Morris and W\"atterstam). Meanwhile, the
  bound states wavefunctions, which describe the 
  physics of the ZZ D--brane (or flux) background in interaction with probe FZZT
  D--branes, are captured by the generalized mKdV system, and
  non--perturbative string equations obtained by reduction of that
  system (the Painlev\'{e} II hierachy found by Periwal and Shevitz in
  this context).

\end{abstract}
\newpage \baselineskip=18pt \setcounter{footnote}{0}

\section{Opening and Closing Remarks}
\label{sec:introduction}
It is safe to say that, at this point in time,  we do not understand
string (or M--) theory as well as we would like. While we have understood and
appreciated that there is a rich bounty of physical phenomena contained
in the theory, this has mostly been uncovered in perturbation theory,
occasionally sweetened by a glimpse into the non--perturbative realm
afforded by special sectors of the theory such as soliton solutions
(including branes of various sorts) or various topological reductions.

The physics that we have so far learned from the theory has provided
numerous promising and exciting phenomenological scenarios that form
the basis for several research endeavours to understand and
incorporate current experimental and observational data from Nature,
and furnish testable predictions about new physics. These endeavours
are still embryonic, and cannot fully mature without much more
understanding of the underlying theory.

Furthermore, much of what we have learned pertains to the critical
string theories, a rich class for study of course, but after all of
the non--perturbative lessons that we have learned in the last decade,
the fact that as a field we mostly still linger in the critical domain
should be regarded as nothing more than the force of habit; so much
historical baggage. Having broken free of the shackles of perturbative
thinking, there is no compelling physical reason to restrict attention
to critical strings in a search for a description of Nature. It is
time to try to move on to other areas of the theory, where the tools
and concepts we need to make contact with Nature may well be waiting
to be found.

There has been some movement. Due to progress in the understanding of
open string sectors in Liouville conformal field
theory\cite{Teschner:2000md,Zamolodchikov:2001ah,Fateev:2000ik}, and
following on from the proposal in ref.\cite{McGreevy:2003kb}, recent
years have seen a growing realisation that the $c$ (or ${\hat c}\leq
1$) non-critical string theories (equivalently $D\leq 2$, so
``subcritical''), despite being rather simple as compared to their
higher dimensional cousins, contain several model examples of the
non--perturbative phenomena that have so fascinated us from higher
dimensional critical strings such as D--branes, holography,
open--closed transitions, tachyon condensation, etc. In fact, this
class of models ---first arrived at by double scaling certain matrix
models\cite{Brezin:1990rb,Douglas:1990ve,Gross:1990vs,Gross:1990aw}---
contains the earliest examples of fully non--perturbative formulations
of string theories, which remain the {\it only} formulations available
where one can ask and answer (appropriate) questions arbitrarily far
from perturbation theory.  Furthermore, the fact that one can get
different string theories by expanding the physics in different small
parameters (something we'd like to better understand about M--theory
and the critical string theories) is manifest in these models. For
example, in one class of models first found and studied extensively in
refs.\cite{Dalley:1992qg,Dalley:1992vr,Dalley:1992yi,Dalley:1992br},
identified as ${\hat c}<1$ type 0A strings in
refs.\cite{Klebanov:2003wg}, and to be further discussed at length in
this paper, the physics is contained rather succinctly in a
non--linear differential equation, with no reference to strings and
their world--sheets. It is only when a small dimensionless parameter
is identified and the solution is expanded in terms of this parameter
does the physics take on the interpretation of a string theory (where
the small parameter is the string coupling) which can be open or
closed depending upon which parameter is taken to be
small\footnote{For some of the $D=2$ strings in this class, there has
  even been a recent attempt to formulate a sort of M--theory
  specifically. See ref.\cite{Horava:2005tt}.}.

The celebrated non--perturbative phenomena mentioned  near the
beginning of the previous paragraph are examples of exciting physics
of which we would like even more examples, and of which we would like
better understanding. The type of non--perturbative formulations under
discussion furnish such examples and enhance our understanding
somewhat by sharpening the terms in which the phenomena of interest are
expressed and by confirming them as robust (perhaps even generic)
non--perturbative features of the theory.

Double scaled matrix models (and their accompanying physics) were
abandoned as non--perturb-ative approaches by the field only a few
years after their first construction, the main reasons cited being
non--perturbative ambiguities (see e.g., discussion and analysis in
refs.\cite{David:1991sk,Shenker:1990uf,David:1990ge}) and
oversimplicity. This was despite clear demonstrations that there were
fully consistent and unambiguous
models\cite{Dalley:1992qg,Dalley:1992vr,Johnson:1992pu,Johnson:1992wr,Johnson:1992uy}
available which avoided these objections, and non--perturbative maps
between models with closed and open strings\cite{Dalley:1992br}. We
should be careful to not make the same mistake twice and again turn our
attention away from these models prematurely. There is an important
question to ask: Now that we have recognised that these models
describe so many of our favourite important non--perturbative
phenomena, can we learn from them about new non--perturbative physics
that has hitherto been overlooked?

It is with this question in mind that we continue our investigations
in this area. In this paper we explore further a number of the
observations reported in our previous paper\cite{Carlisle:2005mk}. We
sharpen the observations and explicitly extend several features.  A
particular theme which has arisen, and that is quite striking in the
results of this paper and the last is the fact that the underlying
connections to structures from certain integrable systems seem to be
becoming physically clearer and broader in scope.  For example, while
it has been known for some time\cite{Gross:1990aw,Douglas:1990dd} that
the Korteweg--de Vries (KdV) hierarchy of flows organises the closed
string operator content in these models, it was only recently realised
(in our previous paper\cite{Carlisle:2005mk}) that the well--known
B\"acklund transformations that change the number of solitons in a
solution of the KdV equation actually have meaning in this context:
They change the number of background ``ZZ''\cite{Zamolodchikov:2001ah}
D--branes in the model, in one regime, or the number of units of
background R--R flux in another.  That there is a one--to--one
correspondence between D--branes (usually thought of as solitons in a
very different sense) and solitons of KdV (the prototype solitons) is
both ironic and interesting.  We will explore this and the role of the
B\"acklund transformation further in this paper. Closely allied to the
B\"acklund transformation is the Miura map which connects solutions of
the KdV hierarchy to that of another integrable hierarchy, the
``modified'' KdV (mKdV) system. It was noticed long ago in
ref.\cite{Johnson:1992pu} that this map is invertible when applied to
solutions of the string equations for (what we now know is) the type
0A system, and connects them to solutions of another set of equations,
the Painlev\'e II hierarchy.  These equations were known (from the
work of Periwal and Shevitz\cite{Periwal:1990gf,Periwal:1990qb}) to
arise from double scaling unitary matrix models, but it was not
understood exactly what was their string theory interpretation.  The
work of ref.\cite{Johnson:1992pu} therefore gave an interpretation,
for the first time, for the double scaled unitary matrix models: They
were really secretly systems of open and closed strings, just written
in rather apparently strange variables defined by the Miura map. Now
we know that there is more, and in this paper we make explicit the
role of the unitary matrix models and their associated Painlev\'e II
string equations. The physics derived from those systems is simply
that of a probe
``FZZT''\cite{Fateev:2000ik,Teschner:2000md,Zamolodchikov:2001ah}
D--brane when it has been stretched entirely in the Liouville
direction, terminating on a family of ZZ D--branes living at positive
infinity. The worldsheet expansions obtained from the solution of the
Painlev\'e II equation give all of the open string worldsheets
associated to this configuration.

We show that the partition functions of this special probe
configuration are actually a type of non--normalizable threshold bound
state wavefunction of an associated supersymmetric quantum mechanics
(a system well--known to be associated to the KdV--mKdV hierarchies),
and their non--normalizability serves to spontaneously break the
supersymmetry. These particular wavefunctions are a special case of
the Baker--Akheizer function of the integrable system, already
identified\cite{Klebanov:2003wg} as being the partition function of
FZZT D--brane probes.

This sharper understanding of the interconnected role of the KdV,
mKdV, B\"acklund, Miura, Baker--Akheizer structures in terms of
familiar physical objects in open and closed string theory is rather
pleasing, and highly suggestive.

\section{A Quantum Mechanics Problem and a String Theory}
\label{sec:setting}
\begin{figure}[ht]
\begin{center}
\includegraphics[scale=0.22]{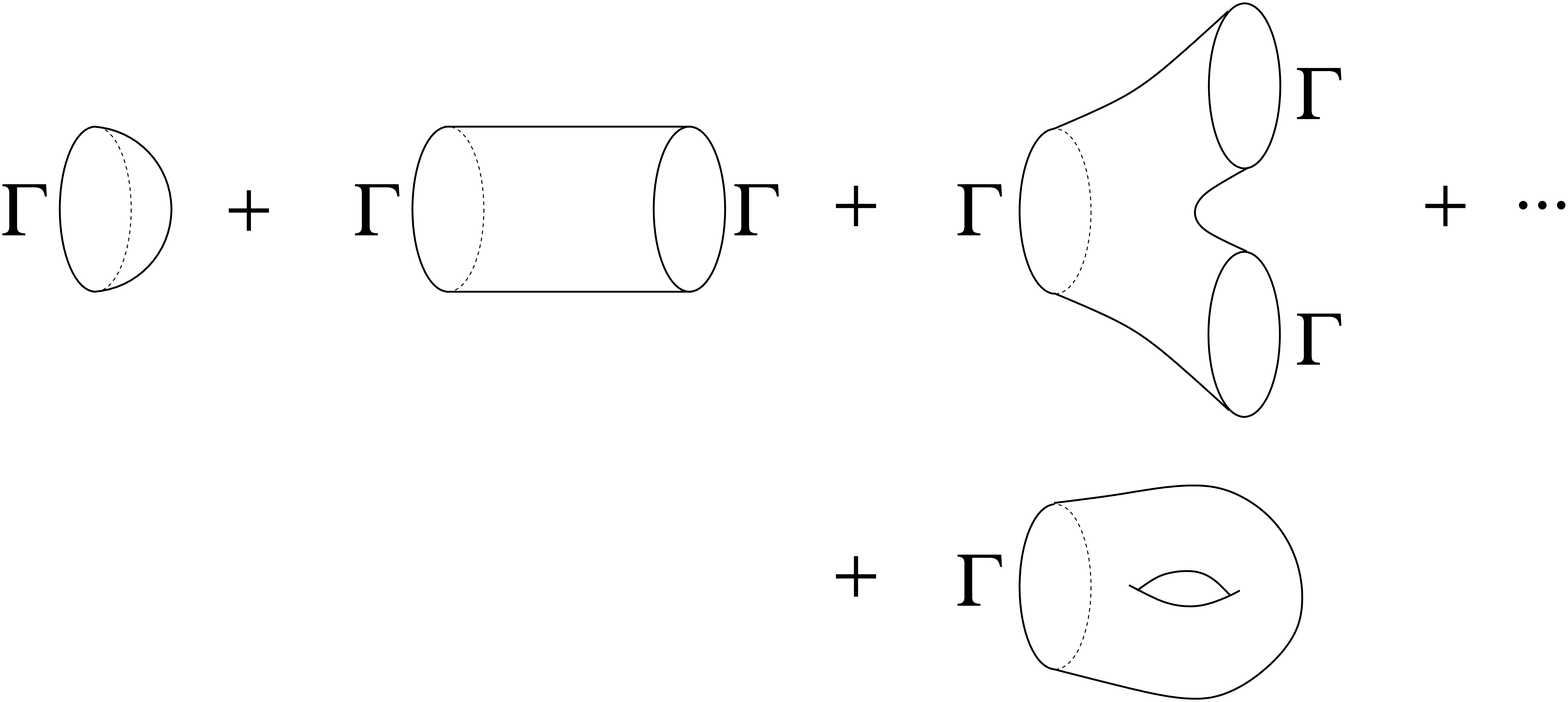}  
\end{center}
\caption{\small Some diagrams contributing to the worldsheet expansion \reef{eq:forwardfree} given by perturbatively solving the string equation for large positive $z$.} 
\label{fig:zzbranes}
\end{figure}

Consider the following one--dimensional quantum mechanics problem:
\begin{equation}
{\cal H}\psi(z)=\lambda\psi(z)\ ,
  \labell{eq:quantummechanics}
\end{equation}
where the Hamiltonian is:
\begin{equation}
  {\cal H}=-Q=-\nu^2\frac{\partial^2}{\partial z^2}+u(z)=-d^2+u\ .
\labell{eq:hamiltonian}  
\end{equation}
The potential $u(z)$ satisfies the differential equation\cite{Dalley:1992br}:
\begin{equation}
u{\cal R}^2-\frac{1}{2}{\cal R}{\cal R}^{''}+\frac{1}{4}({\cal
R}^{'})^2
  =\nu^2\Gamma^2\ .\labell{eq:nonpert}
\end{equation}
and here ${\cal R}=u(z)-z$. Our physics will require that $u(z)$ is a
real function of the real variable~$z$.  A prime denotes $\nu
\partial/\partial z$, and $\Gamma$ and $\nu$ are constants. The
physics also imposes the following boundary conditions:
\begin{eqnarray}
u&\to& z+O(\nu);\qquad \mbox{as}\quad z\to+\infty\nonumber\\
u&\to& 0+O(\nu^2);\qquad \mbox{as}\quad z\to-\infty\ .
  \labell{eq:classical}
\end{eqnarray}
Evidently $\nu$ plays the role of $\hbar$ in this system. The
classical limit ($\hbar\to0$) will be the classical limit of the
underlying string theory, as we shall see: Taking a few more terms in
the expansion of $u(z)$, we have:
\begin{eqnarray}
u&\to& z+\frac{\nu\Gamma}{z^{1/2}}-\frac{\nu^2\Gamma^2}{2z^2}+\frac{5}{32}\frac{\nu^3}{z^{7/2}}\Gamma(4\Gamma^2+1)\cdots;\qquad \mbox{as}\quad z\to+\infty\nonumber\\
u&\to& 0+\frac{\nu^2(4\Gamma^2-1)}{4z^2}+\frac{\nu^4}{8}\frac{(4\Gamma^2-1)(4\Gamma^2-9)}{z^5}\cdots;\qquad \mbox{as}\quad z\to-\infty\ .
  \labell{eq:expansion} 
\end{eqnarray}
The  partition function $Z=\exp(-F)$ of the
string theory is obtained from $u(z)$ using:
\begin{equation}
u(z)=\nu^2\frac{\partial^2 F}{\partial \mu^2}\Biggl|_{\mu=z}\ ,
  \labell{eq:partfun}
\end{equation}
where $\mu$ is the coefficient of the lowest dimension operator in
the world--sheet theory. So integrating twice, we get an expansion:
\begin{eqnarray}
F&=&\frac{1}{6}g_s^{-2}+ \frac{4}{3}\Gamma g_s^{-1}+\frac{1}{2}\Gamma^2 g_s^0\ln\mu+\frac{1}{24}\Gamma(4\Gamma^2+1)g_s^1+\cdots;\qquad \mbox{as}\quad z\to+\infty\labell{eq:forwardfree}\\
F&=&-\left(\Gamma^2-\frac{1}{4}\right)g_s^0\log\mu+\frac{1}{96}(4\Gamma^2-1)(4\Gamma^2-9)g_s^2+\cdots;\qquad \mbox{as}\quad z\to-\infty\ ,
  \labell{eq:freeexpand}
\end{eqnarray}
which is an asymptotic expansion in the dimensionless string coupling
$g_s={\nu/\mu^{3/2}}$. In this expansion, since the sphere term is
non--universal and can be dropped (since it contains no non--analytic
behaviour in $\mu$), the positive $z$ region is a purely open string
expansion, and $\Gamma$, which comes multiplying every worldsheet
boundary, has the interpretation as counting the number of species of
D--brane present\cite{Dalley:1992br}. These are fully localized ``ZZ''
D--branes. See figure~\ref{fig:zzbranes}.

\begin{figure}[ht]
\begin{center}
\includegraphics[scale=0.22]{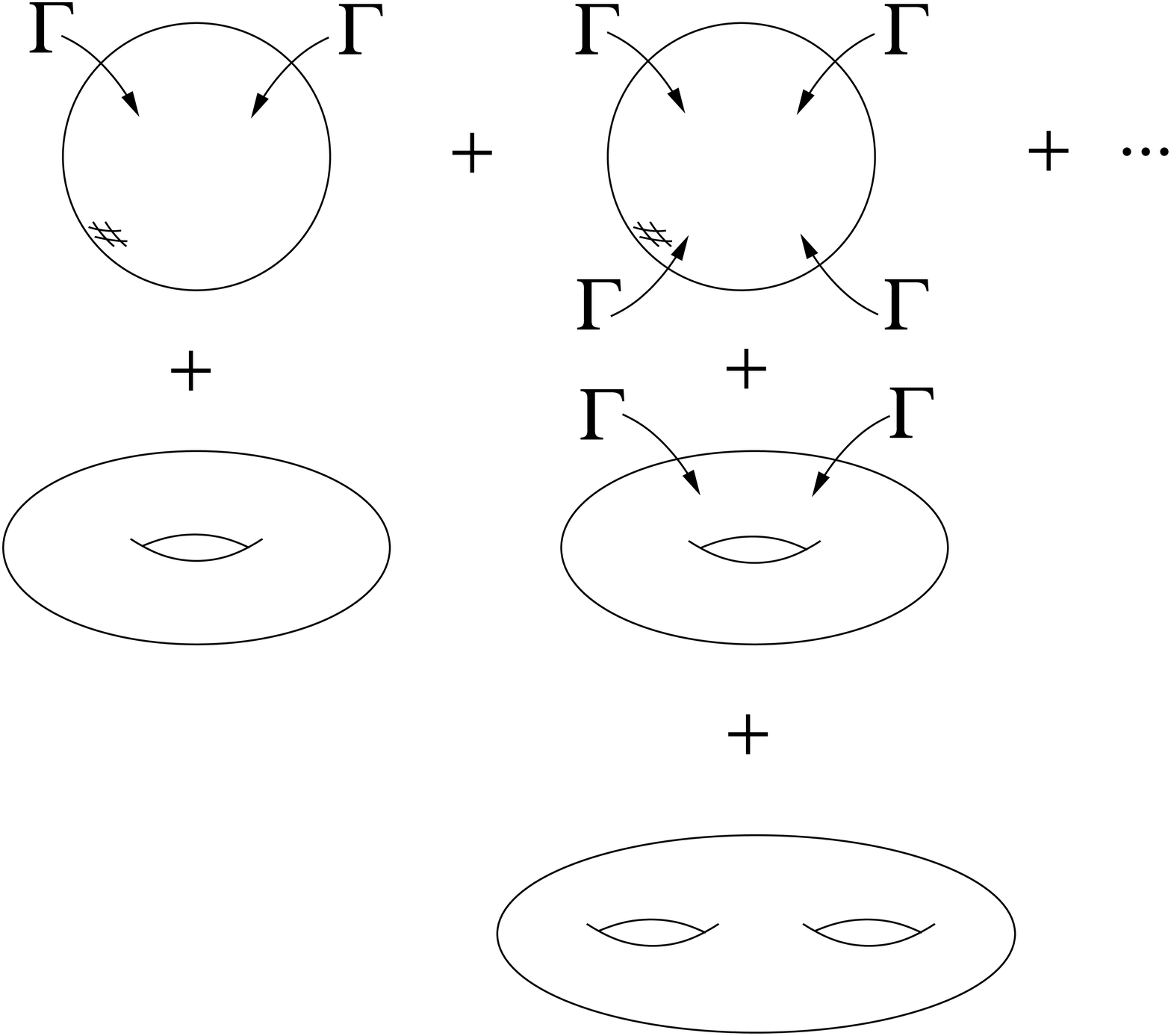}  
\end{center}
\caption{\small Some diagrams contributing to the worldsheet expansion \reef{eq:freeexpand} given by perturbatively solving the string equation for large negative $z$. Each vertex operator insertion gives a factor $g_s\Gamma$. There must be an even number of insertions.} 
\label{fig:flux}
\end{figure}

The negative $z$ region is a purely closed string expansion, and
$\Gamma g_s$ accompanies a worldsheet insertion of a vertex operator
in the presence of $\Gamma$ units of background R--R flux. See
figure~\ref{fig:flux}.  Each vertex operator insertion gives a factor
$g_s\Gamma$. There must be an even number of insertions, as shown by a
worldsheet computation in the continuum theory\cite{Klebanov:2003wg}.

\begin{figure}[ht]
\begin{center}
\includegraphics[scale=0.55]{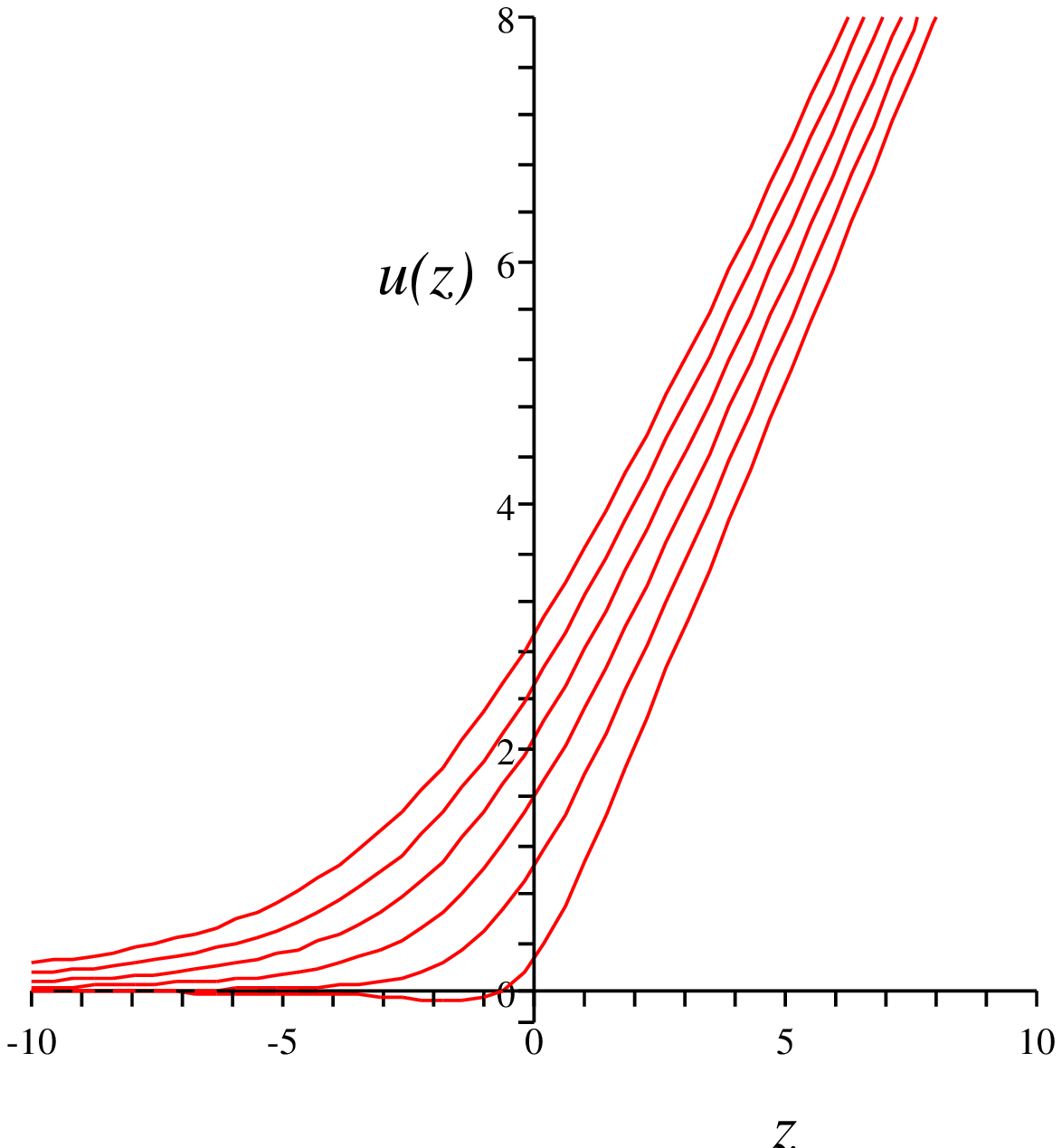} \includegraphics[scale=0.55]{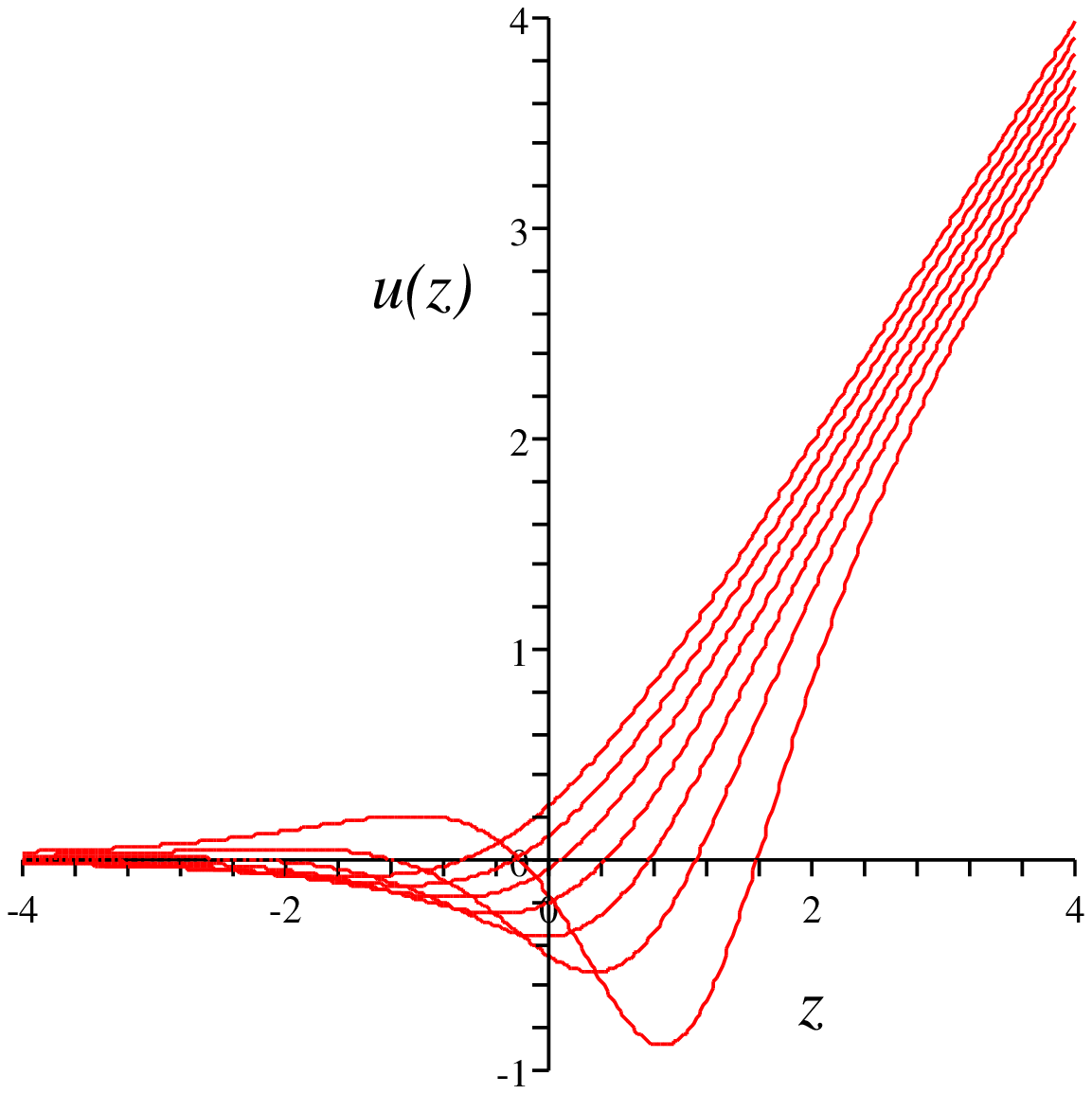} 
\end{center}
\caption{\small Numerical solutions to equation~\reef{eq:nonpert}
for $u(z)$: {\it (a)} cases of positive integer $\Gamma$,
 and {\it (b)} some  cases of $-1<\Gamma<0$.} 
\label{fig:gammaplots}
\end{figure}

\begin{figure}[ht]
\begin{center}
\includegraphics[scale=0.65]{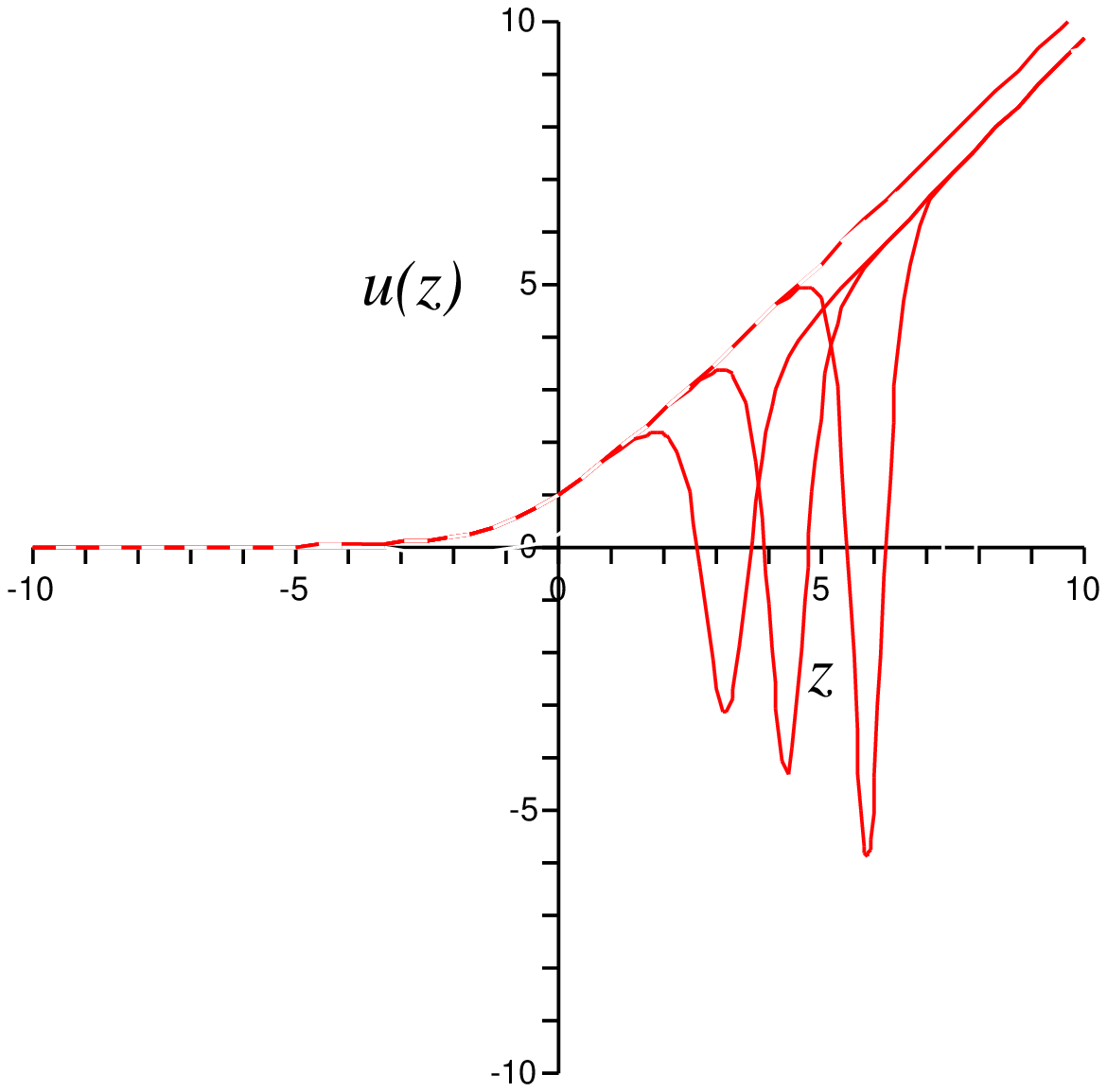} 
\end{center}
\caption{\small Three numerical examples $u(z)$ of $\Gamma=-1+\epsilon$ for $\epsilon$ small and positive, generated by using the B\"acklund transformations on $\Gamma=\epsilon$ curves. They increasingly approach the $\Gamma=+1$ curve (shown dotted at the top), in a manner which is consistent with a limit where the exact identity $u(z)_{\Gamma=-1}=u(z)_{\Gamma=+1}$ is satisfied (see text).} 
\label{fig:moregammaplots}
\end{figure}

Let us look at the potential. Figure~\ref{fig:gammaplots} shows the
typical features, discussed in detail in ref.\cite{Carlisle:2005mk}.
In figure~\ref{fig:gammaplots}{\it (a)} we show only the case of
positive integer $\Gamma$, while in figure~\ref{fig:gammaplots}{\it
  (b)} we show the case of $-1<\Gamma<0$. Note that it becomes
progressively more difficult to solve for $u(z)$ with the given
boundary conditions as $\Gamma$ approaches minus one. However the
B\"acklund transformation (defined later in
equation~\reef{eq:Back-Explicit-Gam}) defined below allows us to
overcome this difficulty, and the results are displayed in
figure~\ref{fig:moregammaplots}.

In ref.\cite{Carlisle:2005mk} we showed, using a combination of
numerical and analytic studies, that the case of integer $\Gamma$ is
rather distinct from that of non--integer $\Gamma$. Furthermore, we
showed there that once one restricts to the integers, the positive
integers are selected by the system, since there is a
non--perturbative transformation which relates $u_\Gamma$ and
$u_{\Gamma\pm1}$ (which was first deduced in ref.\cite{Dalley:1992br},
and made explicit in ref.\cite{Carlisle:2005mk}):
\begin{eqnarray} 
u_{\Gamma \pm 1} = \frac{3 \left(\mathcal{R}^{\prime} \right)^2 -
2 \mathcal{R}\mathcal{R}^{\prime \prime} \mp 8 \nu\Gamma \,
\mathcal{R}^{\prime} + 4 \nu^2\Gamma^2}{4 \mathcal{R}^2}\ , 
\labell{eq:Back-Explicit-Gam}
\end{eqnarray}
where $\mathcal{R} \equiv \mathcal{R}(u_{\Gamma})$, and starting with
$u_{\Gamma=0}$ it is easy to see that $u_{\Gamma=1}=u_{\Gamma=-1}$, as
is suggested\cite{Carlisle:2005mk} in the numerical behaviour of the
function for arbitrary $\Gamma$ studies (see
figure~\ref{fig:moregammaplots}). By extension, it is clear that
$u_\Gamma=u_{-\Gamma}$\ . This is consistent with the theory being
charge conjugation invariant. In fact, one can run the argument the
other way around: Starting with a requirement that the theory be
charge conjugation invariant, the properties of the equations and
their solutions that we uncovered are enough to prove that $\Gamma$
must be a positive integer.  It was also noticed in
ref.~\cite{Carlisle:2005mk} that the transformation above is in fact
the celebrated B\"acklund transformation of the KdV system,
specialized to our system of solutions.

It was also realized in ref.\cite{Carlisle:2005mk} that $\Gamma$
represents the formal soliton number of the solution given by $u(z)$,
further reinforcing the idea that the positive integers are natural
values for it to take. This is why the B\"acklund transformation
naturally acts here. They are well--known to change by an integer the
number of solitons in a KdV solution. We use the term ``formal'' above
because $u(z)$ does not have standard soliton boundary conditions (it
does not vanish in both asymptotic directions), and because the bound
states to which the solitons correspond ({\it via} inverse scattering
theory) are of zero energy ($\lambda=0$ in
equation~\reef{eq:quantummechanics}).

Let us now develop this much further. Turning to study the properties
of our wavefunctions of this system, we can make the following
observations. The potential becomes linear in the large positive $z$
regime, $u(z)=z+\ldots$, and so the wavefunction will behave like an
Airy function in this regime:
\begin{equation}
\psi(z)=(z+\lambda)^{-\frac14}e^{\pm\frac{2}{3}(z+\lambda)^{\frac{3}{2}}}+\cdots
  \labell{eq:airy}
\end{equation}
Meanwhile, in the large negative $z$ regime, the potential vanishes to
leading order, and at next order is:
\begin{equation}
u(z)=\frac{\nu^2}{z^2}\left(\Gamma^2-\frac14\right)+\cdots
  \labell{eq:unextorder}
\end{equation}
so the Schr\"odinger equation is:
\begin{equation}
-\nu^2\frac{d^2\psi(z)}{dz^2}+\frac{\nu^2}{z^2}
\left(\Gamma^2-\frac14\right)\psi(z)=\lambda\psi(z)\ .
  \labell{eq:schro}
\end{equation}
Notice what happens when we change variables using
$\psi(z)=z^{1/2}\phi(z)$, and define  $x=\lambda^{1/2}z/\nu$,
We get the equation:
\begin{equation}
x^2\frac{d^2\phi(x)}{dx^2}+x\frac{d\phi(x)}{dx}+\left(x^2-\Gamma^2\right)\phi(x)=0\ ,
  \labell{eq:besselsequation}
\end{equation}
which is simply Bessel's equation, whose natural solutions are the
Bessel functions, and we choose the modified variety
$\phi(x)=I_\Gamma(x)$, since $\psi(z)$ should be real. 
 The behaviour (for any $\Gamma$) in this regime is, after
converting back to the original variables is:
\begin{equation}
\psi(z)=e^{\frac{\lambda^{\frac12}}{\nu}z}+\cdots
  \labell{eq:asymptoticbessel}
\end{equation}

We wish to focus on the case of vanishing energy, $\lambda$, which is associated with the B\"acklund transformation between 
$u_\Gamma$ and $u_{\Gamma \pm 1}$. This will lead to an understanding of how $\psi(z)$ encodes the physics of our background
D--branes and fluxes. To proceed we factorize the Hamiltonian as:
\begin{equation}
\left({\cal H}_{\Gamma}-\lambda\right)\psi=\left(-d^2+u_\Gamma-\lambda\right)\psi=\left[\left(-d+v\right)\left(d+v\right)-\lambda\right]\psi=0\ ,
  \labell{eq:factorize}
\end{equation}
where
\begin{equation}
  \labell{eq:miura}
  v^2-v^\prime=u_\Gamma\ ,
\end{equation}
The continuum begins at
$\lambda=0$, corresponding to threshold bound states of zero
energy. Specializing to that case, we have $\psi^\prime=-v\psi$, and so we
deduce that (up to a normalization) we can always write our
wavefunctions as:
\begin{equation}
\psi=e^{-\int\! v/\nu}\ ,
  \labell{eq:waveform}
\end{equation}
and ask what equation $v$ satisfies. Some algebra shows that if
$u_\Gamma$, satisfies~\reef{eq:nonpert},
then $v$ satisfies:
\begin{equation}
  \labell{eq:painleveII}
  \frac{1}{2}v^{\prime\prime}-v^3+zv+\nu C=0\  ,
 \quad\mbox{\rm where}\quad C=\frac{1}{2}\pm\Gamma\ .
\end{equation}
This is the Painlev\'e II equation. Actually, there are two
choices for the constant $C$, for every $\Gamma$, which we'll
sometimes write as $C_+$ and $C_-$, although we will sometimes also
use the fact that given either choice for $C$, the other is $1-C$.
This gives two cases of interest for every choice of $\Gamma$, and so
we shall sometimes call these $v_\Gamma$ and ${\bar v}_\Gamma$,
respectively:
\begin{eqnarray}
v_\Gamma&\to&  z^{1/2}+\frac{1}{2}\frac{\nu C_+}{z}-\frac{1}{32}\frac{\nu^2}{z^{5/2}}(12\Gamma^2+12 \Gamma+5)+\cdots\ ; \qquad\mbox{as}\quad z\to 
+ \infty\nonumber\\
v_\Gamma&\to& -\frac{\nu C_+}{z}-\frac{1}{8}\frac{\nu^3}{z^4}(4\Gamma^2-1)(2\Gamma+3)+\cdots\ ; \qquad\mbox{as}\quad z\to - \infty\nonumber\\
{\bar v}_\Gamma&\to& - z^{1/2}+\frac{1}{2}\frac{\nu C_-}{z}+\frac{1}{32}\frac{\nu^2}{z^{5/2}}(12\Gamma^2-12 \Gamma+5)+\cdots\ ; \qquad\mbox{as}\quad z\to 
+ \infty\nonumber\\
{\bar v}_\Gamma&\to& -\frac{\nu C_-}{z}+\frac{1}{8}\frac{\nu^3}{z^4}(4\Gamma^2-1)(2\Gamma-3)+\cdots\ ; \qquad\mbox{as}\quad z\to - \infty\ ,
  \labell{eq:formsofv}
\end{eqnarray}
and so integrating and exponentiating to form the wavefunction we find
that there are two choices for wavefunctions:
\begin{eqnarray}
\psi_\Gamma&\to&  z^{-\frac12(\frac12+\Gamma)}e^{+\frac23 z^{3/2}}+\cdots\ ; \qquad\mbox{as}\quad z\to 
+ \infty\nonumber\\
\psi_\Gamma&\to& z^{-\frac12-\Gamma}+\cdots\ ; \qquad\mbox{as}\quad z\to - \infty\nonumber\\
{\bar \psi}_\Gamma&\to&  z^{\frac12(\frac12-\Gamma)}e^{-\frac23 z^{3/2}}+\cdots\ ; \qquad\mbox{as}\quad z\to 
+ \infty\nonumber\\
{\bar \psi}_\Gamma&\to& z^{-\frac12+\Gamma}+\cdots\ ; \qquad\mbox{as}\quad z\to - \infty\ ,
  \labell{eq:formsofvii}
\end{eqnarray}
This is appropriate, since the potential $u(z)\to z$ as $z\to+\infty$
and so the wavefunction should resemble the exponential tail of the
Airy function in that limit, and it does. Meanwhile in the
$z\to-\infty$ limit, where the potential vanishes to leading order,
the wavefunction has a purely power law behaviour, entirely
appropriate for a zero energy state. Rather than have $\Gamma$
different wavefunctions for the $\Gamma$ different objects of
degenerate energy (which one ought not to expect in one dimensional
quantum mechanics) $\psi(z)$ and ${\bar \psi}(z)$ should be thought of
as describing $\Gamma$ bound objects, signalled by the
non--exponential part of their behaviour as $z\to\pm\infty$.  Shortly,
we will further unpack the meaning of $\psi(z)$ and ${\bar \psi}(z)$.

\begin{figure}[ht]
\begin{center}
\includegraphics[clip,scale=0.30]{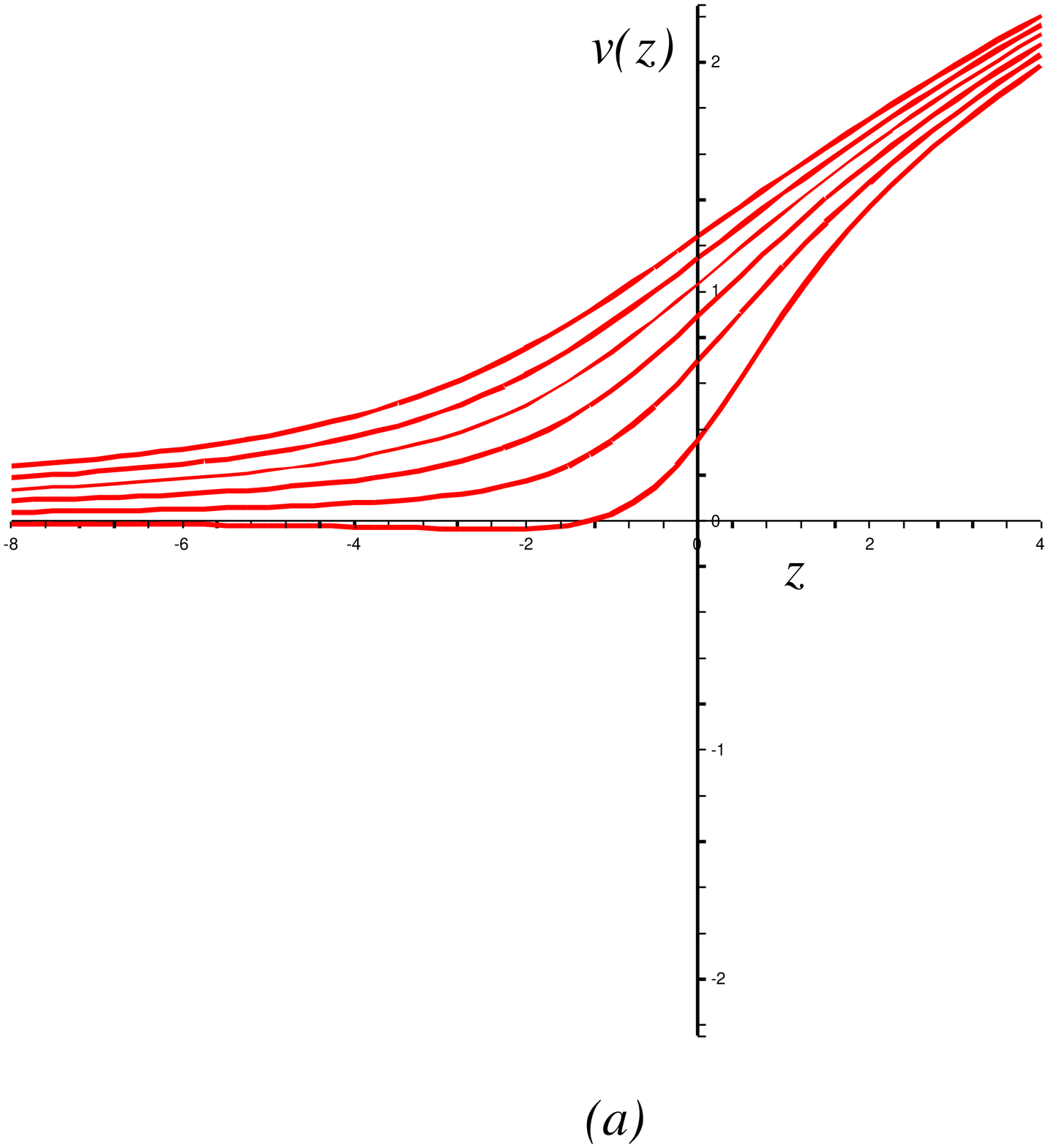}
\includegraphics[clip,scale=0.30]{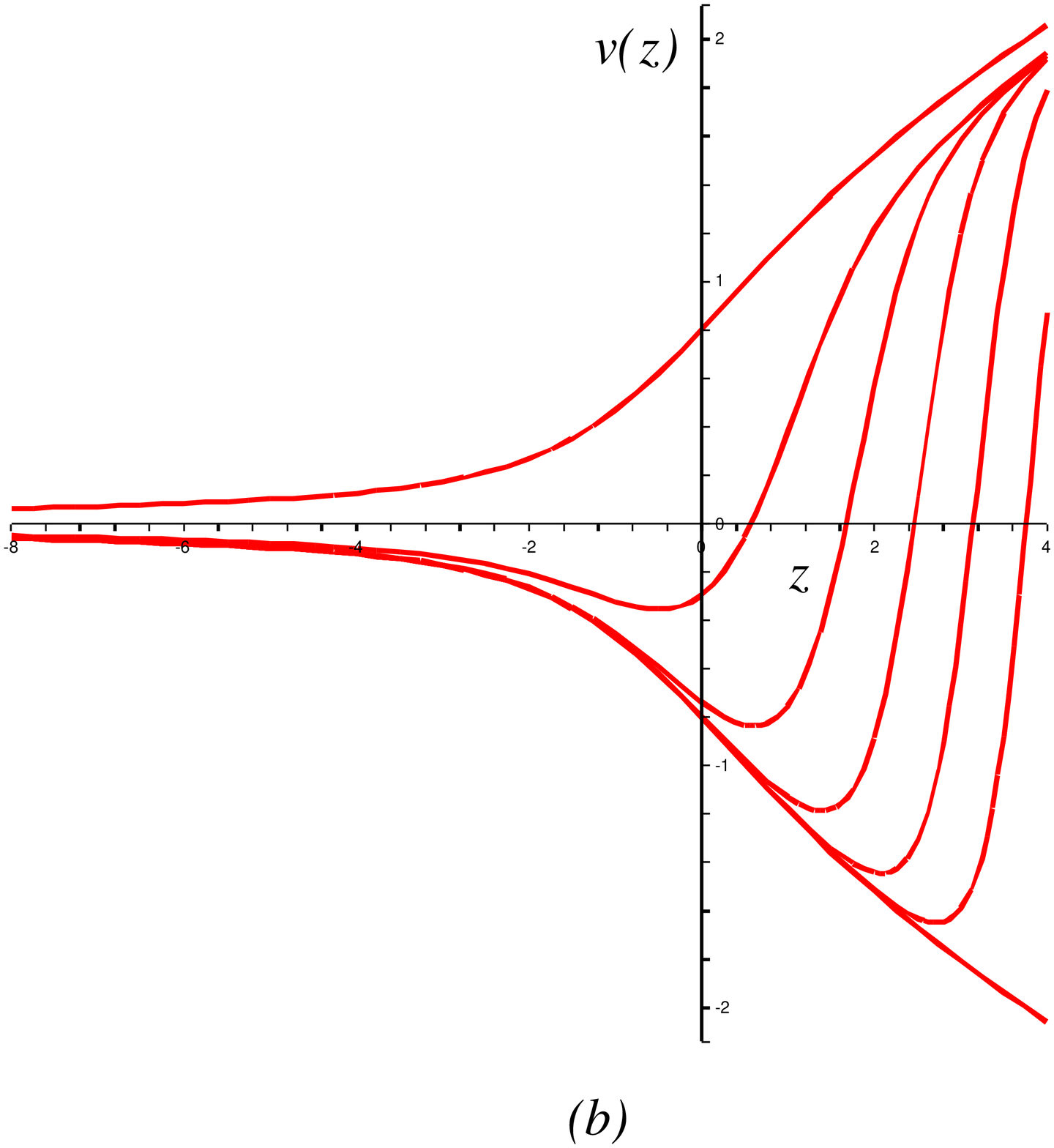}
\end{center}
\caption{\small Numerical solutions for $v(z)$ for a range of values of $\Gamma$. From   top to bottom, we have   {\it (a)}: $\Gamma= 1.4,  1.0, 0.6,  0.2, -0.2,  -0.6 $, and {\it (b)}: $\Gamma=0.0, -0.9, -0.99, -0.999, -0.9999, -0.99999$ generated by the B\"acklund transformations \reef{eq:vBackExplicit}, ending with the curve of ${\bar v}(z)$ for $\Gamma=1$ (see later in text for explanation).} 
\label{fig:veeplots}
\end{figure}

\begin{figure}[ht]
\begin{center}
  \includegraphics[clip,scale=0.30]{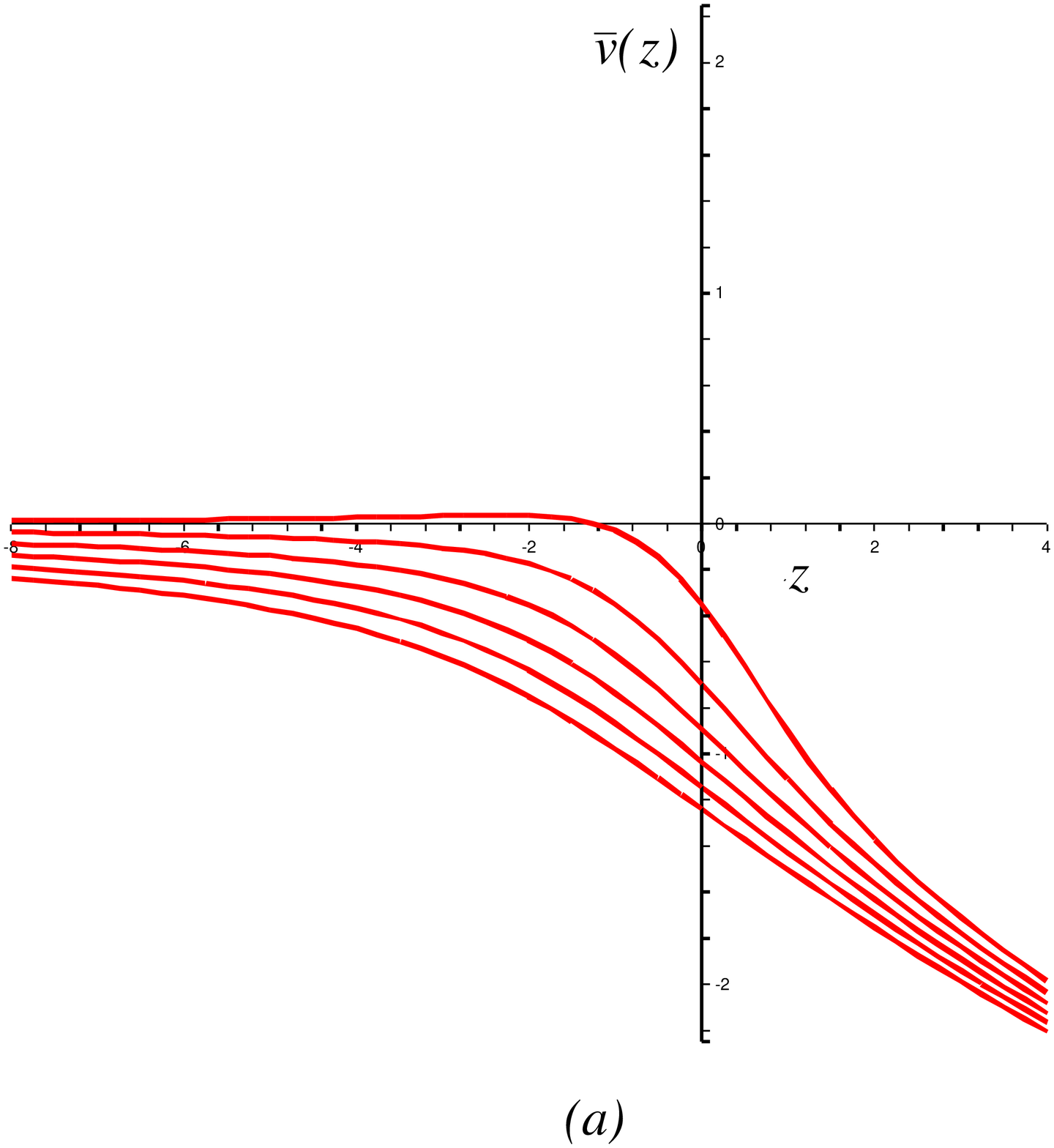}
  \includegraphics[clip,scale=0.30]{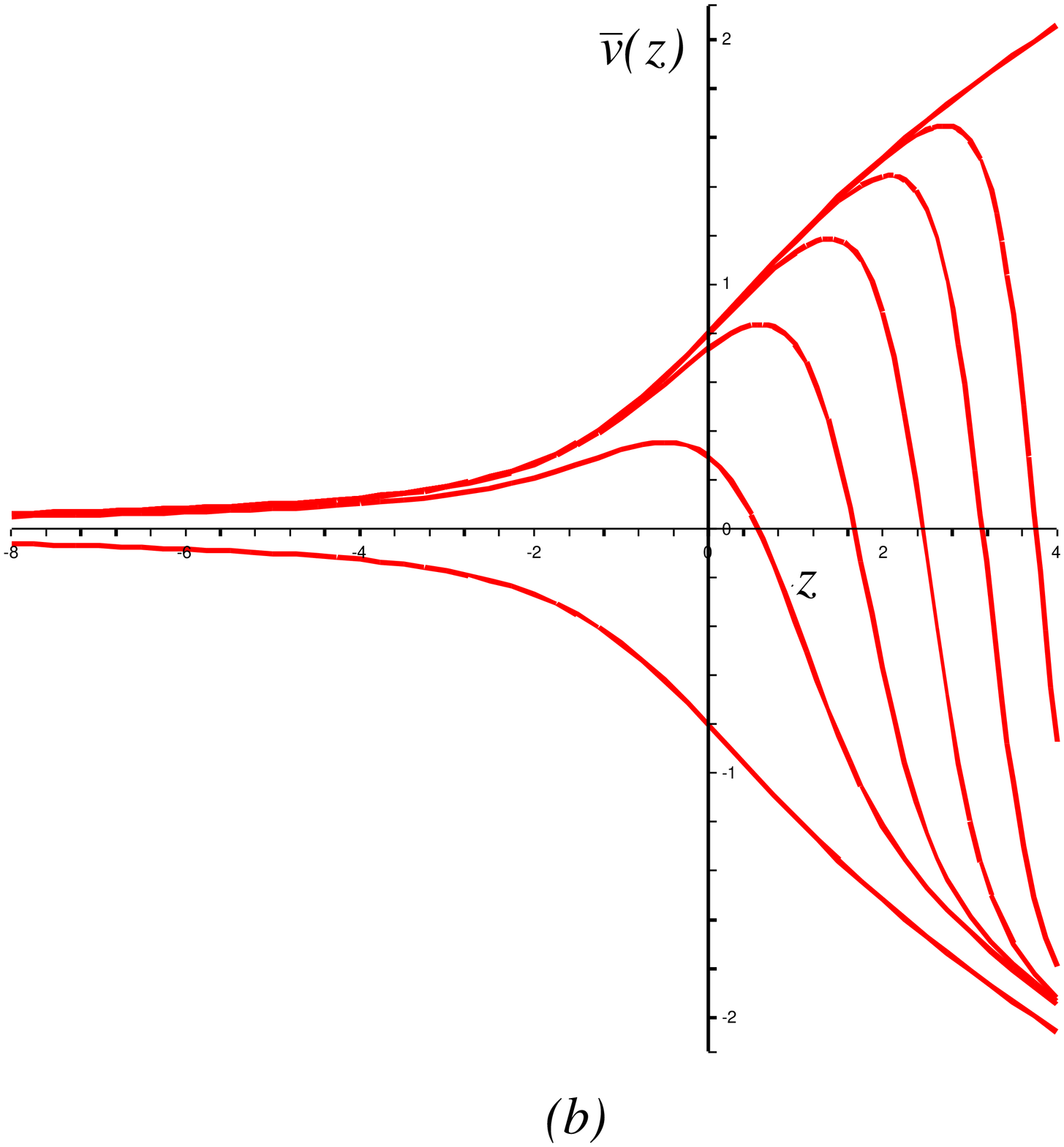}
\end{center}
\caption{\small Numerical solutions for ${\bar v}(z)$ for a range of values of $\Gamma$. From bottom to top, we have {\it (a)}: $\Gamma=2.4, 2.0, 1.6, 1.2,  0.8, 0.4$, and {\it (b)}: $\Gamma=1.0, 0.1, 0.01, 0.001, 0.0001, 0.00001$ generated by the B\"acklund transformations \reef{eq:vBackExplicit}, ending with the curve of $v(z)$ for $\Gamma=0$ (see later in text for explanation).} 
\label{fig:veebarplots}
\end{figure}

\begin{figure}[ht]
\begin{center}
\includegraphics[clip,scale=0.30]{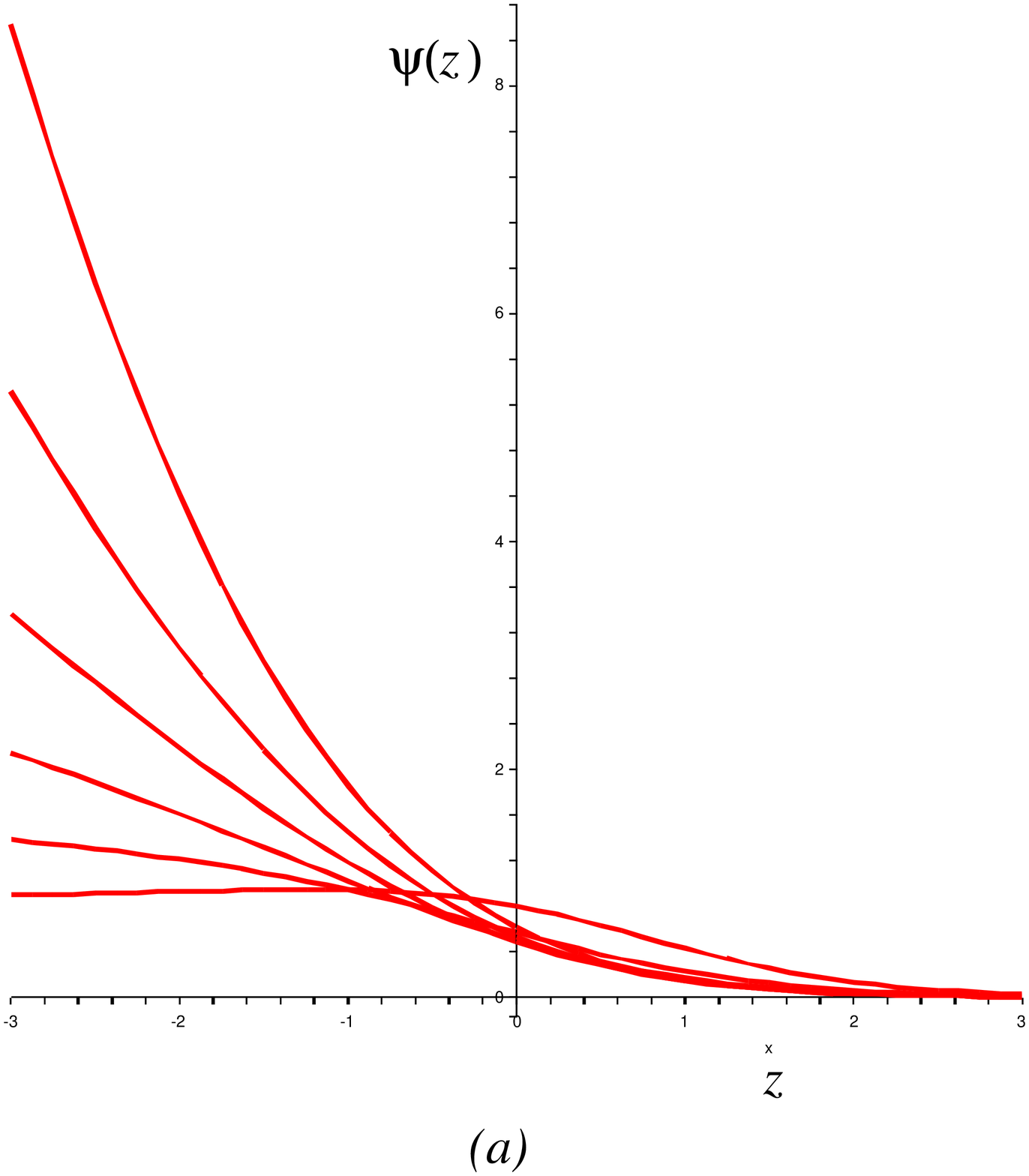}
\includegraphics[clip,scale=0.30]{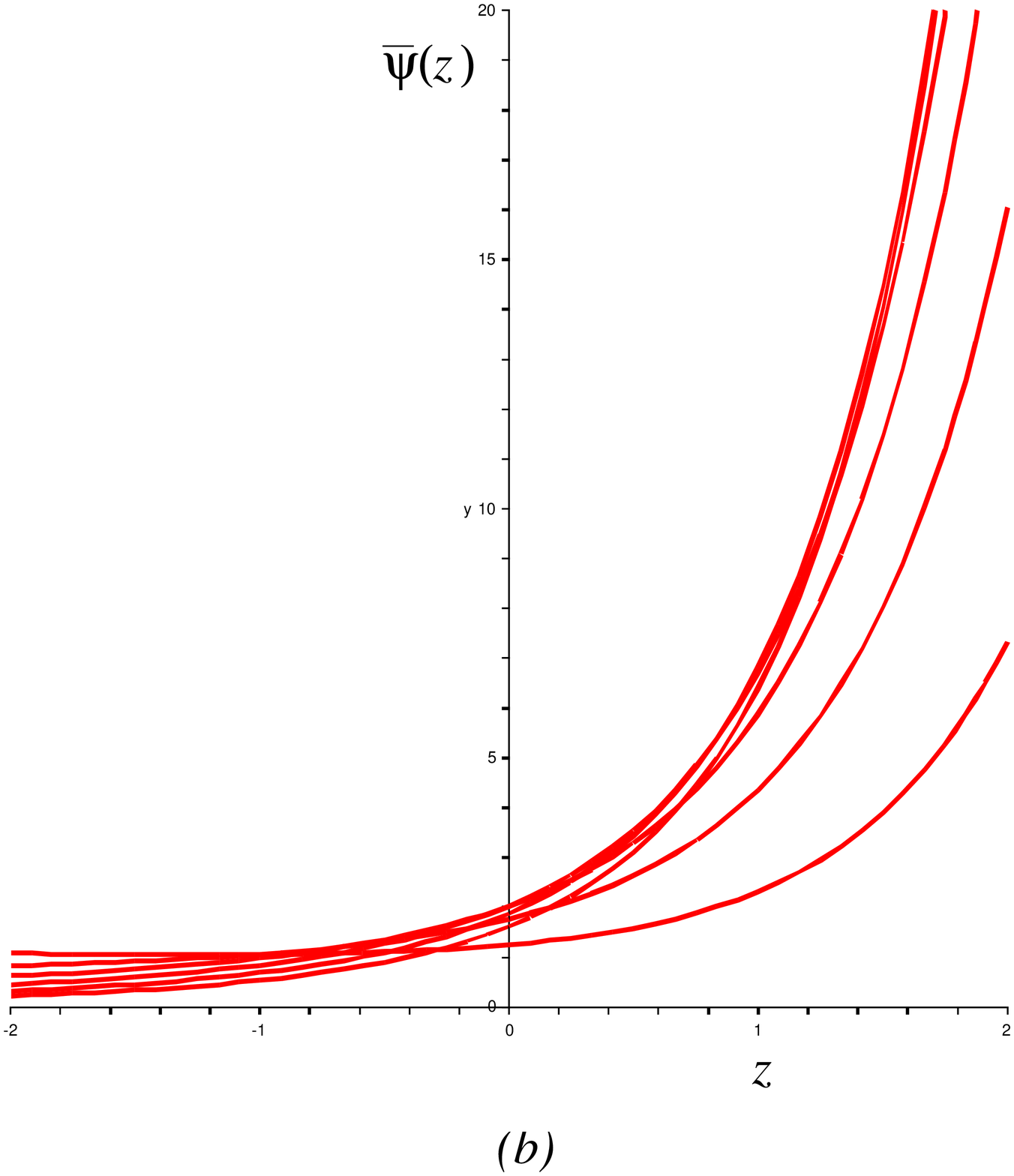}
\end{center}
\caption{\small Numerical solutions for the wavefunctions $\psi(z)$ and ${\bar \psi}(z)$ for a range of values of $\Gamma$ {(\it a)}: Top to bottom, $\Gamma= 1.4,  1.0, 0.6,  0.2, -0.2,  -0.6$; {\it (b)}: Top to bottom, $\Gamma=2.4, 2.0, 1.6, 1.2,  0.8, 0.4$} 
\label{fig:psiplots}
\end{figure}

The form of $v_\Gamma(z)$, ${\bar v}_\Gamma(z)$ and the associated
wavefunctions $\psi_\Gamma(z)$ and ${\bar\psi}_\Gamma(z)$ (recovered
{\it via} equation~\reef{eq:waveform}) can be exhibited numerically,
and examples are given in figure~\ref{fig:veeplots} (for
$v(z)$),~\ref{fig:veebarplots} (for ${\bar v}(z)$), and
figure~\ref{fig:psiplots}, (for $\psi(z)$ and ${\bar \psi}(z)$). Note that it becomes progressively more difficult to solve for $v(z)$ ($\bar{v}(z)$) with the given boundary conditions as $\Gamma$ approaches minus one (zero). However, we will later define B\"acklund transformations \reef{eq:vBackExplicit} for $v(z)$ and $\bar{v}(z)$ that allow us to overcome this difficulty. We have made use of these transformations in figures~\ref{fig:veeplots}{\it (b)} and~\ref{fig:veebarplots}{\it (b)}.

Notice that none of the wavefunctions are normalizable, regardless of
the value of $\Gamma$.  Note also that the case $\Gamma=0$ has $\psi
\sim z^{-1/2}$ to the left and $\psi\sim
z^{1/4}\exp(\pm\frac23z^{3/2})$, so we get a logarithmically divergent
result for its integrated square. In a sense, this wavefunction is just
on the cusp of normalizability. Finally, note that for
$\Gamma=-\frac12$, the wavefunction decays as $\psi(z)\sim
z^\frac12e^{-\frac23 z^{2/3}}$ for large positive $z$, and
$\psi(z)\sim {\rm constant}$ for large negative $z$ with similar
behaviour for ${\bar\psi}(z)$ when $\Gamma=+\frac12$.

\section{Supersymmetry and B\"acklund Transformations}
The factorization process of the previous section
\begin{equation}
{\cal H}_\Gamma={\cal A}^\dagger{\cal A}=-d^2+u_\Gamma\ , 
  \labell{eq:factor}
\end{equation}
where ${\cal A}=d+v$ and ${\cal A}^\dagger=-d+v$ is the first step
in constructing a supersymmetric structure. A ``superpartner''
Hamiltonian can be constructed for ${\cal H}_{\Gamma+1}$ by simply
reversing the order of the factors, to form:
\begin{equation}
{\cal H}_{\Gamma+1}={\cal A}{\cal A}^\dagger=-d^2+u_{\Gamma+1}\ \ , 
  \labell{eq:factoragain}
\end{equation}
where we have changed the label on ${\cal H}$ to reflect the fact that
we now have  a new potential, which is:
\begin{equation}
 u_{\Gamma+1}=v^2+v^\prime\ .
  \labell{eq:newpotential}
\end{equation}
It is easy to see that $\Gamma$ has increased by unity. It follows
from the fact that the Painlev\'e II equation~\reef{eq:painleveII} is
unchanged under $v\to-v$ and $C\to-C$. Therefore, given a solution $v$
to the equation, one can generate a new solution, ${\tilde v}=-v$ to a
new Painleve~II equation whose constant is ${\tilde C}=-C$ instead of
$C$. By the Miura map~\reef{eq:miura}, this defines a new solution
${\tilde u}$ to the string equation~\reef{eq:nonpert}, but with a
different value of the constant on the right hand side, which we can
call ${\tilde \Gamma}$. We can work out what the value of ${\tilde
  \Gamma}$ is by writing ${\tilde C}=\frac12\pm{\tilde\Gamma}$. Since
$C=\frac12+\Gamma$, ${\tilde C}=-\frac12-\Gamma=\frac12-(\Gamma+1)$.
In other words, ${\tilde u}=u_{\Gamma+1}$, and the sign flip on
Painlev\'e~II allows us to construct $u_{\Gamma+1}$ from the ${\bar
  v}_{\Gamma+1}$ function. One can eliminate $v$ from the above formalism 
to obtain a B\"acklund tranformation relating $u_{\Gamma}$ and $u_{\Gamma+1}$ \cite{Carlisle:2005mk}.

\begin{figure}[ht]
\begin{center}
\includegraphics[scale=0.6]{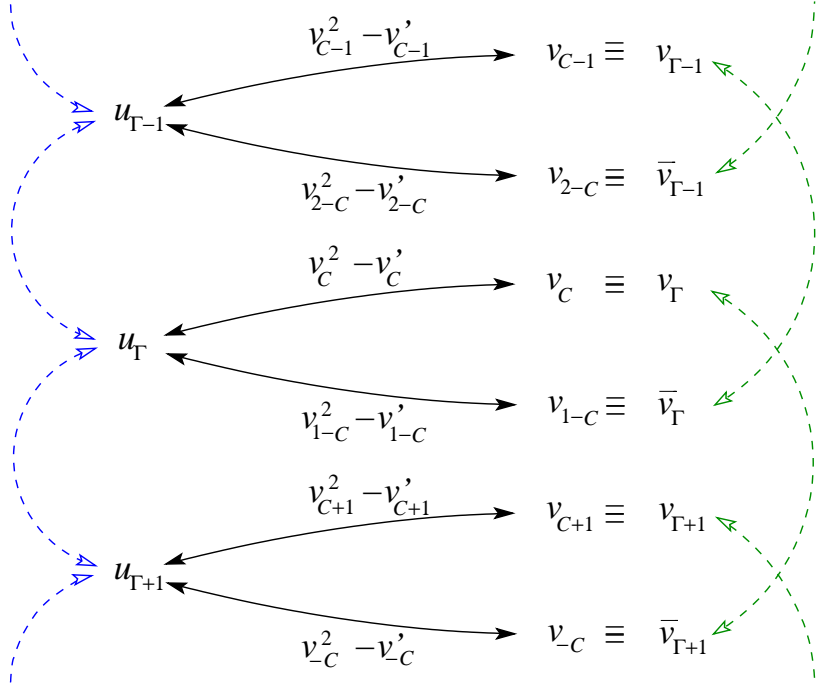}
\end{center}
\caption{\small The structure of how the Miura map in combination with a sign flip on Painlev\'e~II induces the B\"acklund transformations between $u_{\Gamma}$ and $u_{\Gamma\pm1}$.}
\label{fig:backlund}
\end{figure}

It is amusing to note that we can also construct ${\cal H}_{\Gamma-1}$
by instead choosing to do the sign flip on the ${\bar v}_\Gamma$
function. This gives us the equation ${\tilde
  C}=-(1-C)=\frac12+(\Gamma-1)$, telling us that the function ${\tilde
  u}={\bar v}^2+{\bar v}^\prime$ is in fact $u_{\Gamma-1}$, now
constructed from the $v_{\Gamma-1}$   funtion
\begin{equation}
  \labell{eq:backlund}
{\cal H}_{\Gamma-1}=-d^2+{\bar v}^2+{\bar v}\ .   
\end{equation}
In this way we see that we can increase or decrease $\Gamma$ by an
integer depending upon whether we act with the sign flip on $v_\Gamma$
or ${\bar v}_\Gamma$ (respectively), and the resulting $u_{\Gamma+1}$
or $u_{\Gamma-1}$ will be constructed from ${\bar v}_{\Gamma+1}$
and $v_{\Gamma-1}$, respectively. The structure is illustrated in
figure~\ref{fig:backlund}.

Using this re--factorization to construct the Hamiltonian ${\cal
  H}_{\Gamma\pm1}$, we can now note that the spectra of the
Hamiltonians of any two neighbouring~$\Gamma$ are related. If
$\psi(z)$ is a wavefunction of ${\cal H}_\Gamma$ with {\it non--zero}
eigenvalue $\lambda$, then by multiplying the eigenvalue equation on
both sides by ${\cal A}$, one can see that it maps under
``supersymmetry'' to a wavefunction ${\cal A}\psi(z)$ of ${\cal
  H}_{\Gamma\pm1}$ with the {\it same} energy $\lambda$. Away from the
zero energy sector therefore, ${\cal H}_\Gamma$ and ${\cal
  H}_{\Gamma\pm1}$ have identical spectra.  Now there are no states for
$\lambda<0$, (see ref.\cite{Carlisle:2005mk}) and so we learn that the
continuum parts of the spectrum are identical for each Hamiltonian. At
$\lambda=0$, the story is slightly different, however.  There, the map
fails, and we find that as we map the spectrum from ${\cal H}_\Gamma$
to that of ${\cal H}_{\Gamma+1}$, any $\lambda=0$ state is lost.

In the standard nomenclature, we can think of ${\cal H}_{\Gamma}$ as
bosonic and ${\cal H}_{\Gamma\pm1}$ as fermionic, and we have such a
pair for any value of $\Gamma$.  A most efficient way of writing all
of this to see the supersymmetric structure is to define the identity
matrix, $\sigma_0$, together with the Pauli matrices, $\sigma_j$,
$j=1,2,3$:
\begin{equation}
\sigma_0=\begin{pmatrix}1&0\cr0&1\end{pmatrix}\ ,\quad \sigma_1=\begin{pmatrix}0&1\cr1&0\end{pmatrix}\ ,\quad\sigma_2=\begin{pmatrix}0&-i\cr i&0\end{pmatrix}\ ,\quad\sigma_3=\begin{pmatrix}1&0\cr0&-1\end{pmatrix}\ .
  \labell{eq:pauli}
\end{equation}
(where here $i$ is the square root of $-1$) and combine ${\cal
  H}_{\Gamma}$ and ${\cal H}_{\Gamma+1}$ into a larger
Hamiltonian~$H$:
\begin{equation}
H=(-d^2+v^2)\mathbf{\sigma}_0-v^\prime\mathbf{\sigma}_3=\begin{pmatrix}-d^2+v^2-v^\prime&0\cr0&-d^2+v^2+v^\prime\end{pmatrix}\ .
  \labell{eq:bighamilton}
\end{equation}
Defining
$\sigma_{\pm}=\frac{1}{2}(\sigma_1\pm i\sigma_2)$, so that
$\{\sigma_+,\sigma_-\}=1$ and $[\sigma_+,\sigma_-]=-\sigma_3$, we can
define the supercharges as:
\begin{equation}
Q^\dagger=-i{\cal A}^\dagger \sigma_-=i\begin{pmatrix}0&0\cr -d+v&0\end{pmatrix}\ , \qquad  Q=-i{\cal A} \sigma_+=i\begin{pmatrix}0&d+v\cr0&0\end{pmatrix}\ ,
  \labell{eq:supercharges}
\end{equation}
(in the last two displayed equations $v$ can mean $v_\Gamma$ or ${\bar
  v}_\Gamma$, depending upon whether going from ${\cal H}_\Gamma$ to
${\cal H}_{\Gamma+1}$ or to ${\cal H}_{\Gamma-1}$) so that we have the
supersymmetry algebra:
\begin{equation}
\{Q^\dagger,Q\}=2H\  ,\qquad [Q,H]=0\ .
  \labell{eq:susyalgebra}
\end{equation}
The condition for supersymmetry for a particular potential $u(z)$ and
its superpartner potential is that we have a normalizable zero energy
ground state, and we have already seen that for our system's boundary
conditions, this fails. Our family of potentials therefore breaks
supersymmetry.

\section{The Physics of the Wavefunctions}
\label{sec:wavefunctions}
\subsection{The FZZT D-Brane as a Probe}
So we've seen that we have a system with a non--normalizable
wavefunctions, $\psi(z)$, ${\bar \psi}(z)$ for a state of zero energy,
$\lambda$, which we nevertheless keep as part of the physics. The
question naturally arises as to what the meaning of $\psi(z)$ and
${\bar\psi}(z)$ might be, and to what physics its properties described
in the last section correspond.  Somehow it encodes the information
about the presence of $\Gamma$ background objects. As learned in the
context of AdS/CFT, the non--normalizability represents our having
added something to the background: In this case,~$\Gamma$ D--branes,
or units of R--R flux.  How this is encoded in $\psi(z)$ and
${\bar\psi}(z)$ comes initially from expanding $v_{C_\pm}$.  Let us start in
the $z\to+\infty$ limit (recall $v_{C_+}\equiv v_\Gamma$,
$v_{C_-}\equiv {\bar v}_\Gamma$), where from
equation~\reef{eq:formsofv}:
\begin{equation}
v_{C_\pm}=\pm z^{\frac12}+\frac{1}{2}\frac{\left(\frac12\pm\Gamma\right)}{z}\mp\frac{1}{32}\frac{\nu^2}{z^{5/2}}(12\Gamma^2 \pm 12 \Gamma+5)+\cdots
  \labell{eq:expandvee}
\end{equation}
and in constructing the wavefunction~ \reef{eq:waveform}, we are
instructed to integrate once and divide by~$\nu$, which gives:
\begin{eqnarray}
{\cal F}&=&\frac{2}{3}\frac{z^\frac32}{\nu}+\frac{1}{2}\left(\frac12+\Gamma\right)\ln z +\frac{1}{48}\frac{\nu^2}{z^{3/2}}(12\Gamma^2+12 \Gamma+5)+\cdots\nonumber\\
&=&\frac{2}{3}g_s^{-1}+\frac{1}{2}\left(\frac12+\Gamma\right)g_s^0\ln \mu+\frac{1}{48}g_s^1(12\Gamma^2+12 \Gamma+5)+\cdots 
\nonumber\\{\bar {\cal F}}&=&-\frac{2}{3}g_s^{-1}+\frac{1}{2}\left(\frac12-\Gamma\right)g_s^0\ln \mu-\frac{1}{48}g_s^1(12\Gamma^2-12 \Gamma+5)+\cdots 
 \labell{eq:FreeFZZT}
\end{eqnarray}
The rewriting in terms of powers of $g_s=\frac{\nu}{\mu^{3/2}}$ in the
final lines of each equation shows that this is clearly a worldsheet
expansion in surfaces with boundary, but there are {\it two types} of
boundary. Those coming from strings ending on ZZ D--branes have a
factor of $\Gamma$ associated to them (as in the analogous $u(z)$
expansion~\reef{eq:freeexpand}), but there is another type of
boundary. It is {\it not} of ZZ D--brane type, and so it has no factor
of $\Gamma$ associated to them. Such boundaries are present in every
worldsheet.  They are to be associated to a {\it single} FZZT
D--brane, which is in the presence of the background as a probe. This
sum of connected diagrams (all with a boundary on the FZZT D--brane)
is evidently the free energy of the FZZT D--brane in the presence of
the $\Gamma$ background ZZ D--branes. Exponentiation to form the
wavefunction $\psi$ is then the construction of the {\it partition
  function} of this system\cite{Maldacena:2004sn,Seiberg:2004ei}.

\begin{figure}[ht]
\begin{center}
\includegraphics[scale=0.22]{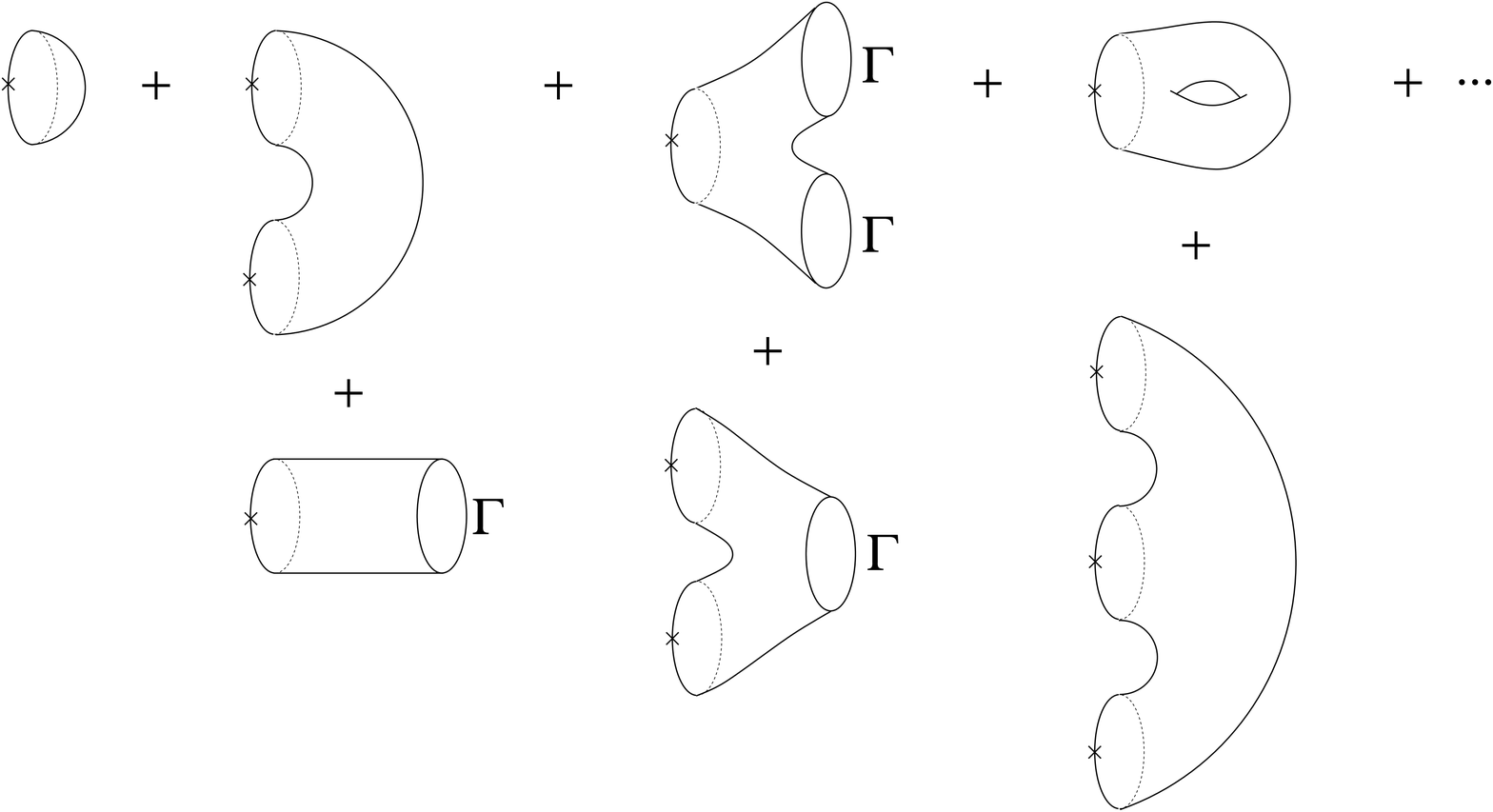}  
\end{center}
\caption{\small Some diagrams contributing from the worldsheet expansion \reef{eq:FreeFZZT} given by perturbatively solving the Painlev\'e~II equation. for large positive $z$. There is a single FZZT D--brane probing a background of $\Gamma$ ZZ D--branes.} 
\label{fig:fzztprobe}
\end{figure}

There is also particular geometrical meaning to the point $\lambda=0$
which we are studying here. The space of scaled eigenvalues, forming a
continuum, $\lambda\in[0,\infty)$, is the natural space that arises
from the underlying double scaled matrix model. Its connection to the
target space of the minimal string theory was emphasized in
refs.\cite{Seiberg:2003nm,Maldacena:2004sn}. The minimal string's most
natural coordinate (in the continuum approach) is the Liouville
direction, $\varphi$, which runs from $-\infty$ to $+\infty$. There is
a linear dilaton $\Phi\propto\varphi$, and so the string coupling
$g_s=e^\Phi$ grows with increasing~$\varphi$. There are two distinct
types of
D--brane\cite{Fateev:2000ik,Teschner:2000md,Zamolodchikov:2001ah}.
The ZZ D--branes are localized in $\varphi$, but are in the strong
coupling region at $\varphi\to +\infty$. The FZZT D--branes are
extended in $\varphi$, but dissolve and come to an end at a specific
value $\varphi=\varphi_\lambda$. The label $\lambda$ is used because
this value is related to our $\lambda$ as $\varphi\sim-\ln\lambda$. So
$\lambda$ space is the moduli space of FZZT D--brane
positions\cite{Klebanov:2003wg}, and at a given $\lambda$, the
wavefunction $\psi(z)$ is the partition function telling us about the
physics of the FZZT D--branes {\it via} open string worldsheets.  So
the case $\lambda=0$, which we've been studying so far, is the extreme
case of extending the end of the FZZT D--brane probe all the way up to
touch the $\Gamma$ ZZ D--branes residing at $\varphi=+\infty$. This
explains rather nicely the form of the expansion that we obtain from
the Painlev\'e II equation in this situation. We have all possible
diagrams which start on the FZZT branes; ones which can end on the
background D--branes, and ones which do not. See
figure~\ref{fig:fzztprobe}. In particular, the leading diagram is just
the disc diagram measuring the tension of the probe brane as
$\tau_{\rm fzzt}=\frac{2}{3}g_s^{-1}$.

\begin{figure}[ht]
\begin{center}
\includegraphics[scale=0.22]{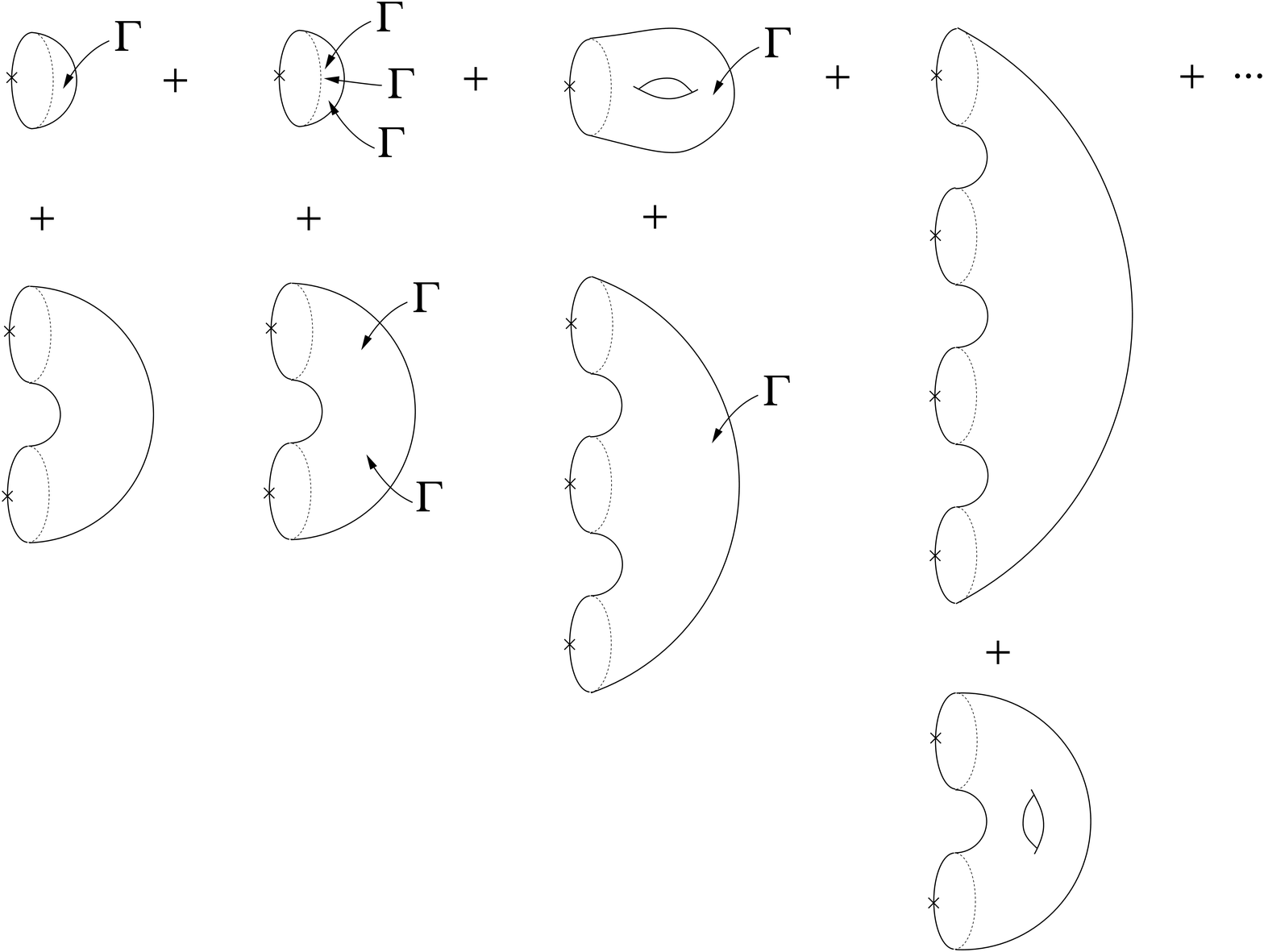}  
\end{center}
\caption{\small Some diagrams contributing from the worldsheet expansion \reef{eq:freeagain2} given by perturbatively solving the Painlev\'e~II equation. for large negative $z$. There is a single FZZT D--brane probing a background with $\Gamma$ units of flux.} 
\label{fig:fzztprobeflux}
\end{figure}

There is also the limit $z\to-\infty$. Expanding $v_{C_\pm}$ in this
limit (see equation~\reef{eq:formsofv}), dividing by $\nu$ and
integrating once gives:
\begin{eqnarray}
{\cal F} 
&=&-\left(\frac12+\Gamma\right)g_s^0\ln \mu+ \frac{1}{24}g_s^2\left(4\Gamma^2-1\right)\left(2\Gamma+3\right)+\cdots \labell{eq:freeagain1}
\end{eqnarray}
\begin{eqnarray}
{\bar {\cal F}}&=&-\left(\frac12-\Gamma\right)g_s^0\ln \mu- \frac{1}{24}g_s^2\left(4\Gamma^2-1\right)\left(2\Gamma-3\right)+\cdots  \labell{eq:freeagain2}
\end{eqnarray}
which, like equation~\reef{eq:FreeFZZT} for the large positive $z$
expansion, has a purely open string explanation. The $\Gamma$ ZZ
D--branes have been replaced by background R--R flux, which the FZZT
D--brane now probes.  The diagrams again all contain one FZZT
boundary, and there are also diagrams with a puncture by a vertex
operator, each such puncture bringing a factor $\Gamma g_s$, as in the
$u(z)$ expansion for the background given in
equation~\reef{eq:freeexpand}.See figure~\ref{fig:fzztprobeflux}.
There is a selection rule that the total number of boundaries and
punctures, taken together, must be even. This presumably has the same
origins as for the similar restriction discussed beneath
equation~\reef{eq:freeexpand}.

\section{Making Sense of Morons}
It is worth remarking here that the appearance of the Painlev\'e~II
equation we've noted here --controlling the physics of the fully
extended FZZT branes-- is different from the manner in which it
appears as the physics of the 0B version of this
model\cite{Klebanov:2003wg}. In that context, one starts with the
string equation~\reef{eq:nonpert} for $u(z)$, and uses the
transformation of Morris\cite{Morris:1991cq} ({\it i.e.} not the Miura
transformation~\reef{eq:miura}):
\begin{equation}
 u(z)=f^2(z)-z\ ,
  \labell{eq:morris}
\end{equation}
to find that {\it if $\Gamma=0$}, then $f(z)$ satisfies a
Painlev\'e~II equation, with $C=0$. If $\Gamma\neq0$, the resulting
equation is quite different, as there is an extra term
$\nu^2\Gamma^2/f^3$. A further difference is that the type~0B free
energy is the second integral of $f^2(z)$, not the first integral of
$f(z)$. So on the one hand, these systems are clearly different, and
the appearance of Painlev\'e~II in both contexts is a coincidence, but
on the other hand, it is an interesting clue as to how to make sense
of the cases of $\Gamma=\pm\frac12$ here\footnote{This is the reason
  for the title of this subsection. Integer~$\Gamma$ counts the number
  of D--branes, which are also instantons of the matrix model. An
  object with half instanton number is called a ``meron'' in field
  theory, and so $\Gamma=\pm\frac12$, whether it be half a brane or
  not, deserves the name ``moron''. (This joke has been waiting in
  reserve since it arose in a 1996 conversation with Atish Dabholkar,
  who thought it up in the context of fractional branes in the
  orientifold models of
  refs.\cite{Gimon:1996rq,Gimon:1996ay,Dabholkar:1996pc}.)}.  The
functions $v_C(z)$ or $f(z)$ do have interpretations in their own
right, the first being a one--point function in the FZZT (with
background ZZ D--branes) system, and the second being the square root
of a two--point function of the type~0B/0A system. So we should take
any relation between them as useful information. When
$\Gamma=\pm\frac{1}{2}$, we can have $C=0$ in our Painlev\'e~II
equation determining $v_C$. In other words, under the Miura
map~\reef{eq:miura}, $v_C$ satisfies Painlev\'e~II with $C=0$, and so
we should therefore use the work of ref.\cite{Klebanov:2003wg} and
interpret this as the type~0B model with no background D--branes or
fluxes!  Furthermore, we can use the Morris map~\reef{eq:morris}, to
interpret this as a type~0A string again, with no background D--branes
or fluxes. So it would seem that there is a role for non--integer
$\Gamma$ after all, but it is subtle: We take $\Gamma=\pm\frac{1}{2}$
here, giving us what would appear to be a poorly behaved 0A/0B theory,
but the sector corresponding to an FZZT brane probing this background
can be expressed in terms of correlation functions of the well-defined
0A/0B system with no background. This surely deserves to be further
explored.

\section{General String Equations and Threshold Wavefunctions}

In the last sections we focused on the case of the simplest type~0A
minimal string model, the $k=1$ case of the $(2,4k)$ superconformal
series coupled to superLiouville. This was in order to attempt to
maintain as clear a presentation of the key facts as possible. Much of
what was stated extends naturally to other $k$, and so it is now worth
stating aspects of the overall structure, for completeness.

\subsection{The String Equations}

The $(2,4k)$ series of the minimal type 0A string theory, in the
presence of background R--R sources, has a non--perturbative
definition {\it via} the following ``string equation'', which has the
structure already shown in equation~\reef{eq:nonpert}, and repeated
here for reference\cite{Dalley:1992br}:
\begin{equation}
u{\cal R}^2-\frac{1}{2}{\cal R}{\cal R}^{''}+\frac{1}{4}({\cal
R}^{'})^2
  =\nu^2\Gamma^2\ .\labell{eq:nonpertii}
\end{equation}
 The quantity $\mathcal{R}$ is defined by:
\begin{equation}
\labell{eq:R} {\cal
  R}=\sum_{k=0}^\infty  \left(k+\frac{1}{2}\right)t_k R_k\ ,
\end{equation}
where the $R_k$ ($k=0,\ldots$)
 are polynomials in $u(z)$ and its $z$--derivatives. They
are related by a recursion relation:
\begin{equation}
  \labell{eq:recursion}
  R^{'}_{k+1}=\frac{1}{4}R^{'''}_k-uR^{'}_k-\frac{1}{2}u^{\prime}R_k\ ,
\end{equation}
and are fixed by the constant $R_0$, and the requirement that the rest
vanish for vanishing $u$. The first few are:
\begin{equation}
  \labell{eq:firstfew}
  R_0=\frac{1}{2}\ ;\quad R_1=-\frac{1}{4}u\ ;\quad R_2=\frac{1}{16}(3u^2-u^{''})\ .
\end{equation}
The $k$th model is chosen by setting all the other $t \,$s to zero
except $t_0\equiv -4z$, and $t_k$, the latter being fixed to a
numerical value such that ${\cal R}={\cal D}_k-z$. The ${\cal
D}_k$ are normalised such that the coefficient of $u^k$ is unity,
{\it e.g.}:
\begin{eqnarray}
{\cal D}_1=u\ ,\quad {\cal D}_2=u^2-\frac{1}{3}u^{''}\ ,\quad {\cal D}_3=u^3- u u^{''}-\frac{1}{2}(u^{'})^2+\frac{1}{10}u^{''''}\ .
  \labell{eq:diffpolys}
\end{eqnarray}

For the $k$th model, equation~\reef{eq:nonpert} has asymptotics:
\begin{eqnarray}
  u(z)&=&z^{\frac{1}{k}}+\frac{\nu\Gamma}{kz^{1-\frac{1}{2k}}}+\cdots\quad \mbox{\rm for}\quad z\longrightarrow +\infty\ ,
\nonumber\\
u(z)&=&\frac{\nu^2(4\Gamma^2-1)}{4z^2}+\cdots\quad \mbox{\rm for}\quad z\longrightarrow -\infty\ .
  \labell{eq:largez}
\end{eqnarray}

The function $u(z)$ defines the partition function $Z=\exp(-F)$ of the
string theory {\it via} the equation~\reef{eq:partfun}, where $\mu$ is
the coefficient of the lowest dimension operator in the world--sheet
theory. The partition function
has perturbative expansions in the dimensionless string coupling
\begin{equation}
  \labell{eq:stringcoupling}
g_s=\frac{\nu}{\mu^{1+\frac{1}{2k}}}\ .
\end{equation}

{}From the point of view of the $k$th theory, the other $t_k$s
represent coupling to closed string operators ${\cal O}_k$. It is
well known that the insertion of each operator can be expressed in
terms of the KdV flows\cite{Douglas:1990dd,Banks:1990df}:
\begin{equation}
  \labell{eq:kdvflows}
  \frac{\partial u}{\partial t_k}= R^{'}_{k+1}\ .
\end{equation}

\subsection{The Wavefunctions and The Painlev\'e II Hierarchy}
In ref.\cite{Dalley:1992br}, it was established that the double scaled
unitary matrix models of refs.\cite{Periwal:1990gf,Periwal:1990qb} had
an interpretation as continuum string theories with open string
sectors. This was done by establishing a direct connection (using the
Miura map~\reef{eq:miura}) between the solutions of the string
equations in equation~\reef{eq:nonpert} and solutions of the string
equations for those unitary matrix models.

That connection was interesting, and while it was a considerable
advance in understanding that double scaled unitary matrix models were
intimately related to closed and open string theories obtained by
double scaling Hermitian and Complex matrix models, and furthermore
represents one of the earliest realizations that open and closed
string physics could be related non--perturbatively, it was still not
clear exactly how the map fit with a lot of the recent work on minimal
strings. For example, as already mentioned, the connection of Complex
matrix model formulations of type~0A to string equations for type~0B
written as deformations of Painlev\'e~II were made in
ref.\cite{Klebanov:2003km} using the Morris map~\reef{eq:morris}. Some
progress in understanding the results of that paper in a modern
context was made in our previous paper\cite{Carlisle:2005mk}, where we
used the map from one system to another and back to derive the map
(the explicit B\"acklund transformation) from $u_\Gamma$ to
$u_{\Gamma\pm1}$. But no explicit role was given to $v_{C_\pm}$ as a natural
object in its own right; it was merely a device to facilitate the
derivation of the $\Gamma$--changing map.

Now we see clearly the role of all of the physics of
ref.\cite{Dalley:1992br}. At $\lambda=0$, the wavefunction $\psi(z)$.
can always be written in the form given in
equation~\reef{eq:waveform}, and the function $v_{C_\pm}$ satisfies the
equations of the Painlev\'e~II hierarchy. These equations are most
naturally written in terms of the quantities $S_k$, where:
\begin{eqnarray} \labell{eq:DJMW-Poly}
S_k \equiv \frac12 R_k^\prime [v^2 - v^\prime] - v R_k[v^2 -
v^\prime]\ .
\end{eqnarray}
The equations  of the Painlev\'e~II hierarchy are
\begin{eqnarray} \labell{eq:DJMW-5}
\sum_{k=1}^{\infty} \left(k + \frac12\right) t_k S_k [v(z)] + z v(z) + \nu C=0\ .
\end{eqnarray}
Let us briefly recall the proof of the map between the equations and
its invertibility.  We define $u(z)$ and $v(z)$ such that:
\begin{eqnarray} \labell{eq:DJMW-7}
X_{\pm} [u, v] \equiv \frac12 \mathcal{R}^\prime [u] \mp \nu\Gamma -
v(z)\mathcal{R}[u] = 0\ ,
\end{eqnarray}
which implies a specific form for $v(z)$ given a $u(z)$:
\begin{eqnarray} \labell{eq:DJMW-8}
v = \frac{\frac12 \mathcal{R}^\prime [u] \mp \nu\Gamma
}{\mathcal{R}[u]}\ .
\end{eqnarray}
Noting the identity\cite{Dalley:1992br}:
\begin{eqnarray} \labell{eq:DJMW-9}
0 = X_{\pm}^2 \pm \nu\Gamma X_{\pm} - \mathcal{R}[u] X_{\pm}^\prime
\equiv (v^2 - v^\prime) \mathcal{R}^2 [u] - \frac12 \mathcal{R}[u]
\mathcal{R}^{\prime \prime}[u] + \frac14 (\mathcal{R}^\prime
[u])^2 - \nu^2\Gamma^2\ ,
\end{eqnarray}
we see that if the inverse transformation $u=v^2 - v^\prime$, Miura
map~\reef{eq:miura} exists, then this is just our original string
equation~\reef{eq:nonpert}. On substitution, equation~\reef{eq:DJMW-8}
gives equation~\reef{eq:DJMW-5} with $C=1/2 \pm \Gamma$.  Since the
unitary matrix models of refs.\cite{Periwal:1990gf,Periwal:1990qb}
were originally derived with $C=0$, those models turn out to be
identified with the cases $\Gamma=\pm1/2$.

\label{sec:conclusion}

Finally, note that the flows between $v(z)$s for different models are
organised by the mKdV hierarchy:
\begin{eqnarray} \labell{eq:DJMW-Flow}
\frac{\partial v}{\partial t_k} = \frac12  S^\prime_{k}[v]\ ,
\end{eqnarray}
which implicitly defines for us flows between wavefunctions of our
system, in other words, defining flows between FZZT D--brane
partition functions of the different models.

\section{The Role of the B\"acklund Transformations}
Recall that the B\"acklund transformation for our string equation's
solutions $u(z)$ can be derived by combining the Miura
map~\reef{eq:miura} with the sign flip symmetry of the Painlev\'e~II
equation~\reef{eq:painleveII}, resulting in, for example:
\begin{eqnarray} \labell{eq:Fred1}
u_{\Gamma+1} = v_{\Gamma}^2 + v_{\Gamma}^\prime\ , \\ 
u_{\Gamma} = v_{\Gamma}^2 - v_{\Gamma}^\prime \ , \labell{eq:Fred2}\\
u_{\Gamma-1} = {\bar v}_{\Gamma}^2 + {\bar v}_{\Gamma}^\prime \ . 
\labell{eq:Fred3}
\end{eqnarray}
Recall that for a given value of $\Gamma$, $u_\Gamma$ is related by
the Miura map to the function $v_{\Gamma}$ (which solves Painlev\'e~II
with a constant $C=\frac12+\Gamma$) and also to ${\bar v}_{\Gamma}$
(which solves Painlev\'e~II with a constant $1-C=\frac12-\Gamma$).
From the discussion under equation~\reef{eq:newpotential}, and as is
clear on the diagram in figure~\ref{fig:backlund}, we also have the
relations:
\begin{eqnarray}
v_\Gamma&=&-{\bar v}_{\Gamma+1}\ ,\labell{eq:parity1}\\
\mbox{and}\quad {\bar v}_\Gamma&=&-{ v}_{\Gamma-1}\ ,  \labell{eq:parity2}
\end{eqnarray}
which shall be useful later.
Subtracting equation~\reef{eq:Fred1} from
\reef{eq:Fred2} yields:
\begin{eqnarray} \labell{eq:Interest1}
u_{\Gamma} + 2 v_{\Gamma}^\prime= u_{\Gamma+1} \  ,
\end{eqnarray}
while the difference of equations~\reef{eq:Fred3} and~\reef{eq:Fred2} yields:
\begin{eqnarray} \labell{eq:Interest2}
u_{\Gamma} + 2 {\bar v}_{\Gamma}^\prime= u_{\Gamma-1} \ .
\end{eqnarray}
It is amusing to see how these exact expressions work on the
worldsheet expansions we have discussed previously
(\reef{eq:expansion} for $u(z)$ and \reef{eq:formsofv} for $v(z)$ and
${\bar v}(z)$), and also on the numerical solutions (see
figures~\ref{fig:veeplots} and~\ref{fig:veebarplots}).




So it is as if adding two ($\lambda=0$) FZZT branes given by $v$ is equivalent to
adding one ZZ brane, while adding two ($\lambda=0$) FZZT branes given by $\bar{v}$
has the effect of removing one ZZ D--brane (or flux units). This is
not strictly correct however, because it neglects the interaction
between the two FZZT branes\cite{Maldacena:2004sn}.  Another way of
looking at things is to use equation~\reef{eq:parity1}
and~\reef{eq:parity2} to rewrite equations~\reef{eq:Interest1},
and~\reef{eq:Interest2} as:
\begin{eqnarray}
u_{\Gamma +1} + \bar{v}^\prime_{\Gamma+1} = u_{\Gamma} + v^\prime_{\Gamma}\ ,
\end{eqnarray}
so the background with $(\Gamma+1)$ ZZ branes (or flux units) and one
$\bar{v}$--FZZT brane is the same as the background with $\Gamma$ ZZ
branes (or flux units) and one $v$--FZZT brane.

Recall from the discussion of supersymmetric quantum mechanics that
the Miura map arises from factorising ${\cal H}$ in the form:
\begin{eqnarray} \labell{eq:WaveAgain}
{\cal H}_\Gamma \psi \equiv (-d^2 + u_\Gamma) \psi = (-d + v)(d + v) \psi  
\end{eqnarray}
We found that a solution of this equation is of course just $\psi =
e^p$, where $p' = -v$, since function $p$ clearly satisfies $(d + v)
e^p = 0$. However, there should also exist another function $q$ also
satisfying ${\cal H}_\Gamma e^q = 0$ but not $(d + v)e^q = 0$.
Writing $e^r = (d + v)e^q$ we see that $r$ satisfies the following
two equations:
\begin{eqnarray} \labell{eq:vBack}
(-d + v) e^r = 0 \\
{\cal H}_{\Gamma+1} e^r \equiv (  d+ v)(-d + v) e^r = (-d^2 + u_{\Gamma+1}) e^r = 0\ .
\end{eqnarray}
Using \reef{eq:vBack} we see that $r'=v$ is a solution.
Therefore we can write:
\begin{eqnarray}
e^r = (d + v)e^q = (d - r')e^q = (q' - r') e^q\ ,
\end{eqnarray}
which implies the following auto-B\"acklund transformation between $q$
and $r$:
\begin{eqnarray} \labell{eq:vBack2}
e^{r+q} = q' - r'\ ,
\end{eqnarray}
where $q'$ and $r'$ are both solutions of the Painleve II equation
differing by unit $\Gamma$. This directly relates solution $v_\Gamma$
to solutions $v_{\Gamma \pm 1}$; and similarly for $\bar{v}$. The
B\"acklund transformation \reef{eq:vBack2} is quite cumbersome to use
in practice, but we can in fact derive a more explicit version using
the same equations that led to the explicit version of the $u$
transformation. We find:
\begin{eqnarray} \labell{eq:vBackExplicit}
v_{\Gamma-1} = - v_\Gamma + \frac{2 \nu \Gamma}{\mathcal{R}[v^2_\Gamma - v^{\prime}_{\Gamma}]} \, , \quad
\bar{v}_{\Gamma-1} = - \bar{v}_\Gamma - \frac{2 \nu (\Gamma - 1)}{\mathcal{R}[\bar{v}^2_\Gamma + \bar{v}^{\prime}_{\Gamma}]} \nonumber \\
v_{\Gamma+1} = - v_\Gamma + \frac{2 \nu (\Gamma + 1)}{\mathcal{R}[v^2_\Gamma + v^{\prime}_{\Gamma}]} \, , \quad
\bar{v}_{\Gamma+1} = - \bar{v}_\Gamma - \frac{2 \nu \Gamma}{\mathcal{R}[\bar{v}^2_\Gamma - \bar{v}^{\prime}_{\Gamma}]}
\end{eqnarray}
Using the fact that $v_0 = \bar{v}_0$ it is easy to show inductively
that $v_\Gamma = \bar{v}_{-\Gamma}$ if and only if $\Gamma$ is an
integer! This is borne out numerically in the same
way\footnote{That is, starting with $\Gamma=\epsilon$, B\"acklund
  transforming to $\Gamma=-1+\epsilon$, and letting $\epsilon
  \rightarrow 0$.} that we showed $u_{\Gamma}=u_{-\Gamma}$ for integer
$\Gamma$ (see figures~\ref{fig:moregammaplots},~\ref{fig:veeplots},
and~\ref{fig:veebarplots}). The procedure is identical to that used in the continuation of
Bessel functions, $J_n$, of non--integer $n$ to those of integer $n$ \cite{Carlisle:2005mk}.  Amusingly, we now observe that it fits
rather nicely with the fact that in section~\ref{sec:setting} we noted
that the wave equation \reef{eq:WaveAgain} in the $z \rightarrow
-\infty$ limit is in fact (after a simple change of variables)
Bessel's equation with $n=\Gamma$. So it is not an analogy for $v(z)$
(in that regime) but the {\it identical} system!
  
These results are of course consistent:
\begin{eqnarray}
u_\Gamma = v^2_\Gamma - v^{\prime}_\Gamma = \bar{v}^2_{\Gamma} - \bar{v}^{\prime}_\Gamma  = v^2_{-\Gamma} - v^\prime_{-\Gamma} = u_{-\Gamma}
\end{eqnarray}
So, in many ways it is as if $v$ is an FZZT brane and $\bar{v}$ is an
anti-FZZT brane! The only way that this is consistent under charge
conjugation $\Gamma \rightarrow -\Gamma$ (under which the physics
should be invariant) is for integer $\Gamma$. This is yet more
evidence that $\Gamma$ is quantized.

An interesting observation is that if one adds $v_\Gamma$ and
$\bar{v}_\Gamma$ together asymptotically then one obtains:
\begin{eqnarray}
v_\Gamma + \bar{v}_\Gamma &=& \frac{\nu}{2 z}-\frac{3 \nu^2 \Gamma}{4{z}^{5/2}}+
\frac{15 \nu^3}{32 z^4} +\frac {3 \nu^4 {\Gamma}^{2}}{2 {z}^{4}} -\frac{\nu^5 \Gamma (420 \, \Gamma^2 + 459)}{z^{11/2}} + \cdots \qquad z \rightarrow \infty \nonumber \\
v_\Gamma + \bar{v}_\Gamma &=&  -\frac{\nu}{z} - \frac{\nu^3 (24 \Gamma^2-3)}{4 z^4} - \frac{\nu^5 (240\Gamma^4 - 504 \Gamma^2 + 111)}{16 z^7} + \cdots \qquad z \rightarrow -\infty 
\end{eqnarray}
This can be interpreted entirely in terms of a background containing
just ZZ branes with no FZZT branes. So it is almost as if a brane and
and an anti-brane have annihilated in some way. We have again ignored
the interaction term between the two branes here, so this
`annihilation' would only be correct if the branes had no interaction
with each other. But it is interesting and worthy of note nonetheless.

\section{Remarks About $\tau$--functions}
Another well--known structure in the theory of integrable systems is
the $\tau$--function. The significance of the $\tau$--function in the
context of the minimal string theories was noticed long ago in
refs.\cite{Dijkgraaf:1991rs,Fukuma:1991jw}, where it was shown that
the KdV flows together with the string equation was equivalent to a
family of Virasoro constraints on the square root of the ``closed''
string theory's (what we should now think of as $\Gamma=0$) partition
function. This square root is the $\tau$--function of the KdV
hierarchy, and we have
\begin{equation}
Z_{\Gamma=0} = \tau_0^2\ ,\qquad L_n\tau_0= 0\ ,\quad n\geq -1\ ,
  \labell{eq:virasoro}
\end{equation}
where the $L_n$ are given in terms of products and derivatives with
respect to the $t_k$, and hence the Virasoro constraints can be thought
of as an infinite family of relationships between correlation
functions of the operators to which the $t_k$ couple.

In fact, there is always a natural {\em pair} of $\tau$--functions for
the KdV/mKdV system in general, and one might wonder where the other
one fits into our story.  Some remarks about this were already made at
the end of ref.\cite{Dalley:1992br}, and using the results of
ref.\cite{Johnson:1994vk}, where the complete structure of the
Virasoro constraints (when open string (and flux) sectors are
included) for $\Gamma\neq0$ were presented, together with the insights
gained with the results of the present paper, we can complete the
story.

The notation that is usually used for the pair is $\tau_0$ and
$\tau_1$, but for out purposes it is probably better to, for a given
value of $\Gamma$, use the notation $\tau_\Gamma$ and
$\tau_{\Gamma+1}$, as will become clear. Then, the partition function
of our full string theory at $\Gamma$ is given by the product:
\begin{equation}
  \labell{eq:partfunagain}
  Z_{u}=\tau_\Gamma\tau_\Gamma\ ,
\end{equation}
while the partition function of the FZZT D--brane probe is the ratio:
\begin{equation}
  \labell{eq:partfunfzzt}
  Z_{\rm probe}=\frac{\tau_{\Gamma+1}}{\tau_\Gamma}\ .
\end{equation}
Let us see how this relates to the functions $u_\Gamma$ and
$v_\Gamma$.  We take the logarithm and the second derivative of the
partition function to obtain $u_\Gamma$, and the first derivative for
$v_\Gamma$:
\begin{eqnarray}
u_\Gamma&=&2\frac{\partial^2}{\partial z^2}\log\tau_\Gamma=2\left(\frac{\tau_\Gamma''}{\tau_\Gamma}-\left(\frac{\tau_\Gamma'}{\tau_\Gamma}\right)^2\right)\ .\nonumber\\
v_\Gamma&=&\frac{\partial}{\partial z}\log\frac{\tau_{\Gamma+1}}{\tau_\Gamma}=\left(\frac{\tau_\Gamma'}{\tau_\Gamma}-\frac{\tau_{\Gamma+1}'}{\tau_{\Gamma+1}}\right)\ .
  \labell{eq:relationsuv}
\end{eqnarray}
A little algebra then shows that the combinations $v_\Gamma^2\pm
v_\Gamma '$ do indeed give the functions $u_{\Gamma}$ and
$u_{\Gamma+1}$, if the following relation holds:
\begin{eqnarray}
\tau_{\Gamma+1}''\tau_\Gamma-2\tau_{\Gamma+1}'\tau_{\Gamma}'+\tau_{\Gamma+1}\tau_{\Gamma}''=0\ .
  \labell{eq:relationholds}
\end{eqnarray}
It is  equivalent to the writing of
another partition function:
\begin{equation}
Z_{\rm v}= \tau_{\Gamma}\tau_{\Gamma+1}\ ,
  \labell{eq:vsquared}
\end{equation}
from which the quantity $v^2_\Gamma$ may be derived from a second
derivative of the associated free energy:
\begin{equation}
v_\Gamma^2=\frac{\partial^2}{\partial z^2}\log \,\tau_{\Gamma}\tau_{\Gamma+1}\ .
  \labell{getvee}
\end{equation}
This latter partition function is just that of the original double
scaled unitary matrix
model\cite{Periwal:1990gf,Periwal:1990qb,Hollowood:1992xq}.

Returning to the matter of the Virasoro constraints, it was noted in
ref.\cite{Dalley:1992br} that in the presence of $\Gamma$, the $L_0$
constraint would be modified by a $\Gamma$--dependent constant. While
the constant was not known, the difference between $L_0$ acting on
$\tau_\Gamma$ and $L_0$ acting on $\tau_{\Gamma+1}$ was expected to be
$\frac{C}{2}=\frac{1}{4}\pm\frac{\Gamma}{2}$. The full structure of he
Virasoro constraints for arbitrary $\Gamma$ was worked out in
ref.\cite{Johnson:1994vk}, and the $L_0$ constraint was shown to be
\begin{equation}
  \labell{eq:newellzero}
\left(L_0-\frac{\Gamma^2}{4}\right)\tau_\Gamma=0\ ,
\end{equation}
which is consistent with the result of ref.\cite{Dalley:1992br}, since
upon considering the $L_0$ constraint for a neighbouring value of
$\Gamma$, we get:
\begin{equation}
\frac{(\Gamma\pm1)^2}{4}-\frac{\Gamma^2}{4}=\frac{1}{4}\pm\frac{\Gamma}{2}=\frac{C}{2}\ .
  \labell{eq:sillysums}
\end{equation}

\section{Beyond the Threshold}
\label{beyond}
Let us now consider the case of $\lambda\neq0$, and examine the
structure of $\psi(z)$ at finite energy. For this it is harder to get
an exact non--linear ordinary differential equation (another
deformation of Painlev\'e~II) for, but a route to the perturbative
expansions is as follows. Start again with the factorized form:
\begin{equation}
{\cal H}_{\Gamma}\psi=\left(-d^2+u_\Gamma\right)\psi=\left[\left(-d+v_C\right)\left(d+v_C\right)\right]\psi=0\ ,
  \labell{eq:factored}
\end{equation}
but rather than thinking of it as the case of $\lambda=0$ (as would
have followed from equation~\reef{eq:factorize}), now ask that
\begin{equation}
  \labell{eq:miuramore}
  v^2_C-v_C^\prime=u_\Gamma+\lambda\ ,
\end{equation}
which is an equation for $v_C(z,\lambda)$, and we take our potential
$u_\Gamma(z)$ to satisfy the same string equation as before. Since we
chose that factored form again, we still recover the wavefunction by
exponentiating as in equation~\reef{eq:waveform}.

\subsection{Large Positive $z$}
Returning to the case $k=1$, we get for example (as $z\to+\infty$):
\begin{eqnarray}
v_C\left( z \right) &=&z^{1/2}+\frac12\,{\frac {{\lambda}}{ z^{1/2}}}+
 \left(\frac14+\frac{\Gamma}{2}\right)\frac{\nu}{z}-\frac18\,{\frac {{\lambda}^{2}}{{z}^{3/2}}}-\frac{\lambda}{4}\left(\Gamma+1\right){\frac {1}{{z}^{2}}}
\nonumber\\
&&\hskip0.5cm +
 \frac{1}{32}\left( 2\,{\lambda}^{3}-5-12\,{\Gamma}^{2}-12\,
\Gamma \right) \frac{1}{z^{5/2}}+\frac{\lambda^2}{16}\left(4+3\Gamma\right)\frac{1}{z^{3}}\nonumber\\
&&\hskip0.5cm+\frac{5}{128}\lambda(8\Gamma^2+16\Gamma+10-\lambda^3)\frac{1}{z^{7/2}}\nonumber\\
&&\hskip0.5cm+\left(\frac{1}{2}\Gamma^3+\frac{3}{4}\Gamma^2+\frac{23}{32}\Gamma+\frac{15}{64}-\frac{1}{4}\lambda^3-\frac{5}{32}\lambda^3\Gamma\right)\frac{\nu^2}{z^4}
+\cdots
\end{eqnarray}
Now without question this expansion is a mess. On the face of it,
there is some difficulty in interpreting these terms in the familiar
language of string perturbation theory. It is in fact possible, and
the result is quite elegant.

A clue to what to expect is the knowledge that $\lambda$ controls two
things. On the one hand it is the coefficient of a boundary length
operator, and on the other hand, it is the distance form the tip of
the FZZT D--brane probe at $\varphi=\varphi_c\sim - \ln\lambda$ to the
$\Gamma$ ZZ D--branes located at $\varphi=\infty$. So our worldsheet
expansion should make sense in these terms.

That $\lambda$ is a boundary operator can be seen from the
fact\cite{Banks:1990df} that the expectation value of loops of length
$\ell$ can be represented in terms of the Hamiltonian as:
\begin{equation}
  w(\ell)=\int dz <z|e^{-\ell {\cal H}} |z>\ ,
 \labell{eq:cont}
\end{equation}
and so an ${\cal H}$--eigenvalue $\lambda$ gives it a dependence
$e^{-\ell\lambda}$. So $\lambda$ couples to an operator on the
boundary which measures loop length. Since it is a boundary operator,
one can expect therefore that it will appear in diagrams with boundary
at a given genus with any number of insertions on that boundary. So at
every order in string perturbation theory, $\lambda$ should control an
infinite number of terms corresponding to summing over all ways the
operator can act on each boundary in that diagram. Armed with that
clue, a re--examination of the expansion above yields a useful
rewriting. For example one infinite set of terms can be written at 
 disc order as follows:
\begin{equation}  
z^{1/2}\left(1+\frac{1}{2}\frac{\lambda}{z}-\frac{1}{8}\left(\frac{\lambda}{z}\right)^2+\frac{1}{16}\left(\frac{\lambda}{z}\right)^3+\cdots\right)= z^{1/2}\left(1+\frac{\lambda}{z}\right)^{1/2}=(z+\lambda)^{1/2}\ .
\labell{eq:resum}
\end{equation}
We see that remarkably, the series of terms can be resummed into a
remarkably simple expression. There is a simple interpretation of this
latter result. The result integrates once to give the disc
contribution to the free energy of the probe D--brane. For this $k=1$
model, to leading order in string perturbation theory, the parameter
$z$ already couples naturally to the boundary length (since in the
Hamiltonian, $u(z)\sim z$, and so this follows from
equation~\reef{eq:cont}), and so switching on $\lambda$ in this case
simply renormalizes $z$ additively. We can write the disc contribution
to the free energy from this, after integrating once and dividing by
$\nu$:
\begin{equation}
F_{(0,1,0)}=\frac23 {\tilde g}_s^{-1}\ ,\qquad \mbox{\rm where}\quad {\tilde g}_s=\frac{\nu}{(z+\lambda)^{\frac32}}\ , 
  \labell{eq:freedisc}
\end{equation}
where we have introduced the notation $F_{(b,f,h)}$ to denote the contribution to the
free energy from a diagram with $f$ FZZT D--branes, $b$ ZZ D--branes
and $h$ handles. This resummation turns out to be precisely what happens to all
diagrams involving pure FZZT boundaries and no loops or ZZ boundaries.
For example, let us look at the case of 3 FZZT boundaries. After some
algebra, the infinite seires of $\lambda$ contributions can be
resummed, with the result:
\begin{equation}
F_{(0,3,0)}\sim \frac{\nu}{(z+\lambda)^{\frac{3}{2}}}\sim {\tilde g}_s\ .
  \labell{eq:threestringone}
\end{equation}
The diagram with 1 FZZT D--brane and one handle also gives a
result of the same form, with the total contribution being:
\begin{equation}
F_{(0,3,0)}+F_{(0,1,1)}= \frac{5}{48}{\tilde g}_s\ .
  \labell{eq:threestringandmore}
\end{equation}
Turning to diagrams with both FZZT and ZZ boundaries, let us see what
happens to the three--string vertices. We get, for the case of 2 FZZT
and 1 ZZ, the result:
\begin{equation}
F_{(1,2,0)}=\frac{\nu\Gamma}{4(z+\lambda)z^{1/2}}=\frac14 
{\tilde g}_s\Gamma\left(\frac{z+\lambda}{z}\right)^{\frac12}\ ,
  \labell{eq:threestringtwo}
\end{equation}
while the case of 1 FZZT and 2 ZZ yields:
\begin{equation}
F_{(2,1,0)}=\frac{\nu\Gamma^2}{4(z+\lambda)^{1/2}z}=
\frac{1}{4}{\tilde g}_s\Gamma^2\left(\frac{z+\lambda}{z}\right)\ .
  \labell{eq:threestringthree}
\end{equation}
The pattern emerging is clear. Surfaces have an extra factor of 
\begin{equation}
  \labell{eq:factorextra}
\left(\frac{z+\lambda}{z}\right)^{\frac{1}{2}}\Gamma
\end{equation}
for every boundary that ends on a ZZ D--brane. A $\lambda$--dependent
factor was expected (see above) to appear for every such boundary, in
view of the fact that $\lambda$ sets the separation from the FZZT
D--brane's tip to the $\Gamma$ ZZ D--branes in the background.

Let us consider the structure at higher order in perturbation theory.
The contribution from four FZZT boundaries mixes with that of the
diagram with two FZZT boundaries and one handle to give, after
resumming:
\begin{equation}
F_{(0,4,0)}+F_{(0,2,1)}=\frac{5\nu^2}{128}\frac{d}{dz}\left(\frac{1}{(z+\lambda)^2}\right)=-\frac{5}{64}{\tilde g}_s^2\ .
  \labell{eq:twoplushandleandfour}
\end{equation}
More interesting are the cases mixing the different types of
boundaries. For example, the case of 2 FZZT and 2 ZZ boundaries can be
written as:
\begin{eqnarray}
F_{(2,2,0)}=\frac{\nu^2\Gamma^2}{8}\frac{d}{dz}\left(\frac{1}{(z+\lambda)z}\right)&=&-\frac{\nu^2\Gamma^2}{8}\left[\frac{1}{(z+\lambda)^2z}+\frac{1}{(z+\lambda)z^2}\right]\nonumber\\
&=&-\frac{\nu^2\Gamma^2}{8}\frac{1}{(z+\lambda)^3}\left[\frac{z+\lambda}{z}+\left(\frac{z+\lambda}{z}\right)^2\right]\nonumber\\
&=&-\frac{{\tilde g}_s^2\Gamma^2}{8}\left(\frac{z+\lambda}{z}\right)\left[1+\left(\frac{z+\lambda}{z}\right)\right]\ .
  \labell{eq:fourstringone}
\end{eqnarray}
 
The prefactor shows that this result follows our rule, but there are
two terms which contribute to the diagram. We expect that this relates
to the fact that there are two distinct ways of slicing this diagram
to yield the underlying three--string vertices. Some algebra shows
that the case of 3 ZZ and 1 FZZT yields:
\begin{eqnarray}
F_{(3,1,0)}=\frac{\nu^2\Gamma^3}{12}\frac{d}{dz}\left(\frac{1}{(z+\lambda)^{1/2}z^{3/2}}\right)&=&-\frac{{\tilde g}_s^2\Gamma^3}{24}\left(\frac{z+\lambda}{z}\right)^{\frac32}\left[1+3\left(\frac{z+\lambda}{z}\right)\right]\ ,
  \labell{eq:fourstringtwo}
\end{eqnarray}
which again has the expected prefactor, and that there is a
contribution from two terms fits with the possible decomposition of
the diagram.

At the same order in ${\tilde g}_s$ the results for $F_{(1,3,0)}$ and $F_{(1,1,1)}$ mix to give\begin{equation}
F_{(1,3,0)}+F_{(1,1,1)}=-\frac{5{\tilde g}_s^2\Gamma}{32}\left(\frac{z+\lambda}{z}\right)^{\frac12}\left[1+\frac13\left(\frac{z+\lambda}{z}\right)+\frac15\left(\frac{z+\lambda}{z}\right)^2\right]\ ,
  \labell{eq:fourstringandhandle}
\end{equation}
where now there is a third term, fitting with the fact that there is
one way of slicing the diagram with a handle to decompose it into
three--string vertices.

\subsection{Diagrammatics for Large Positive $z$}
In this section we describe in more detail the general structure of
the resummations uncovered above.  For ease of notation let us define
a symbol:
\begin{eqnarray} \labell{eq:Notation}
(b,f) \equiv z^{-b/2} (z + \lambda)^{-f/2} \ ,
\end{eqnarray}
with $(b,f)^{\prime}$ denoting one differentiation with respect to
$z$, {\it etc.,} and we'll also use $(b,f)^{(k)}$ to denote $k$
differentiations with respect to $z$.  We have
$(b,f)^\prime=-\frac{b}{2}(b+2,f)-\frac{f}{2}(b,f+2)$.  In this
section we will also define the contribution to the free energy from a
particular diagram, $F_{(b,f,h)}$ implicitly to have a factor of
$\nu^{-\chi} \Gamma^{b}$ multiplying it, where $\chi \equiv
2\!-\!2h\!-\!b\!-\!f$ is the Euler number. Once again we will consider
the $v_\Gamma(z)$ expansion; although the $\bar{v}_\Gamma (z)$
expansion can be trivially obtained by multiplying the $F_{(b,f,h)}$
by $(-1)^{f}$.

\begin{figure}[ht]
\begin{center}
\includegraphics[scale=0.22]{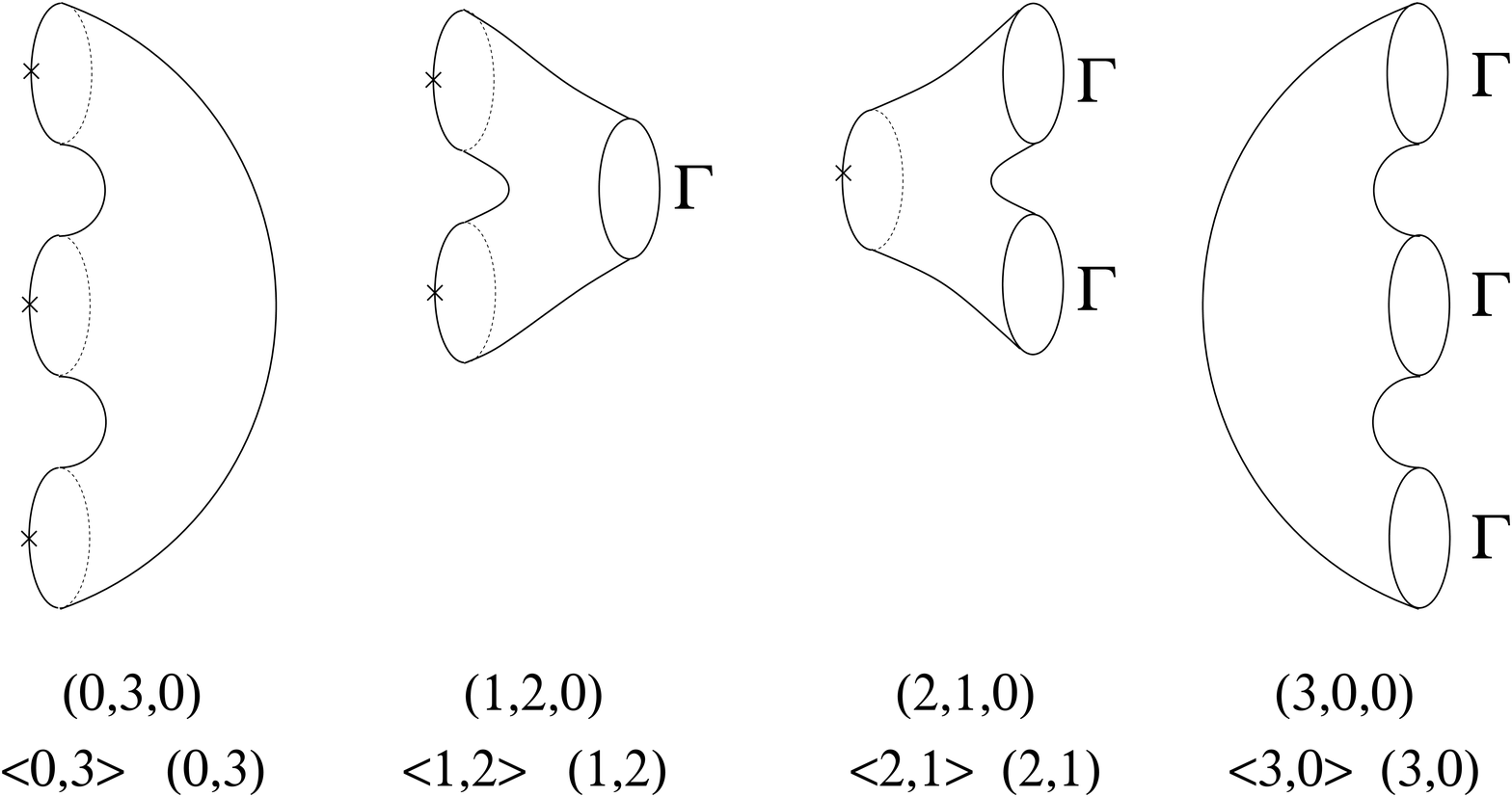}  
\end{center}
\caption{\small Some of the diagrams fundamental to building diagrams of higher topology (and their contributions the free energy) using the methods described in the text. Shown is some of the notation used to describe them in the text in general ($(b,f,h)$), to denote their fundamental status ($\langle b,f\rangle$) and also the shorthand for the terms which appear in the free energy ($(b,f)$, see equation~\reef{eq:Notation}). } 
\label{fig:fundamental}
\end{figure}

For a surface with no handles ($h$=0), we find that its contribution
to the free energy in expansion can be resummed (up to an overall
constant) into the form:
\begin{eqnarray} \labell{eq:GenRel}
F_{(b,f,0)} \propto (b,f)^{(b+f-3)} \equiv \langle b,f \rangle
\end{eqnarray}
If the number of derivatives is negative in \reef{eq:GenRel}, as is
the case with the disc and the cylinder, then one should integrate
instead of differentiating. A similar equation to \reef{eq:GenRel}
holds in the case of $h$ non-zero providing that one or other of $b$
and $f$ is zero instead.  We have:
\begin{eqnarray} \labell{eq:GenRel2}
F_{(b,0,h)} \propto (b,0)^{(b+h-1)} &\equiv& (b+2h,0)^{(b-1)} \ , \quad b,h \neq 0 \nonumber \\ 
F_{(0,f,h)} \propto (0,f)^{(f+h-1)} &\equiv& (0, f+2h)^{(f-1)} \ , \quad f,h \neq 0 \nonumber \\
F_{(0,0,h)} &\propto& (6h-4, 0)^{(-1)}\ .
\end{eqnarray}
When all of $b$, $f$ and $h$ are non-zero simultaneously then matters
become far more complicated. It turns out to be possible to resum
every contribution to the free energy as a sum of terms of the form
$(b,f)$. The number of terms increases with the `complexity' of the
series, and the type of terms present in each resummation follow a
strict and predictable pattern. Below (in
equations~\reef{eq:Resuma},~\reef{eq:Resumb} and~\reef{eq:Resumc}) is
a list of the first several resummed terms of the $v(z)$ expansion
including overall coefficients and written in the standard form of
\reef{eq:GenRel} and~\reef{eq:GenRel2} wherever possible. Recall that
is impossible to distinguish between $F_{(b,f,h)}$ and
$F_{(b,f-2,h+1)}$ in the expansion. That these terms are a mixture of
surfaces can often be seen by the relative lack of simplicity of their
coefficients.
\begin{eqnarray} 
F^{\prime \prime}_{(0,1,0)} &=& \frac{1}{2} (0,1) \, , \quad F^{\prime}_{(0,2,0)} = \frac{1}{4} (0,2) \, , \quad F^{\prime}_{(1,1,0)} = \frac{1}{2} (1,1) \nonumber \\
F_{(0,3,0)} + F_{(0,1,1)} &=& \frac{5}{48} (3,0) \, , \quad F_{(1,2,0)} = \frac{1}{4} (1,2) \, , \quad F_{(2,1,0)} = \frac{1}{4} (2,1) \nonumber \\
F_{(0,4,0)} + F_{(0,2,1)} &=& \frac{5}{128} (0,4)^\prime \, , \quad F_{(1,3,0)} + F_{(1,1,1)} = -\frac{5}{32} (1,5) - \frac{5}{96} (3,3) - \frac{1}{32} (5,1) \nonumber \\
F_{(2,2,0)} &=& \frac{1}{8} (2,2)^\prime \, , \quad F_{(3,1,0)} = \frac{1}{12} (3,1)^\prime 
\labell{eq:Resuma}
\end{eqnarray}
\begin{eqnarray}
 F_{(0,5,0)} + F_{(0,3,1)} + F_{(0,1,2)} &=& \frac{221}{16128} (0,5)^{\prime \prime} \nonumber \\
F_{(1,4,0)} + F_{(1,2,1)} &=& \frac{15}{64} (1,8) + \frac{5}{64} (3,6) + \frac{3}{64} (5,4) + \frac{5}{128} (7,2) \nonumber \\
F_{(2,3,0)} + F_{(2,1,1)} &=& \frac{25}{128} (2,7) + \frac{5}{32} (4,5) + \frac{23}{192} (6,3) + \frac{7}{64} (8,1) \nonumber \\ F_{(3,2,0)} &=& \frac{1}{24} (3,2)^{\prime \prime} \, , \quad F_{(4,1,0)} = \frac{1}{48} (4,1)^{\prime \prime} \nonumber \\ F_{(0,6,0)} + F_{(0,4,1)} + F_{(0,2,2)} &=& \frac{113}{24576} (0,6)^{(3)} 
\labell{eq:Resumb}
\end{eqnarray}
\begin{eqnarray}
F_{(1,5,0)} + F_{(1,3,1)} + F_{(1,1,2)} &=& -\frac{1105}{2048} (1,11) - \frac{1105}{6144} (3,9) - \frac{221}{2048} (5,7) \nonumber \\ &-& \frac{165}{2048} (7,5) - \frac{385}{6144} (9,3) - \frac{105}{2048} (11,1) \nonumber \\
F_{(2,4,0)} + F_{(2,2,1)} &=& -\frac{15}{32} (2,10) -\frac{45}{128} (4,8)-\frac{69}{256} (6,6)-\frac{29}{128} (8,4)-\frac{7}{32} (10,2) \nonumber \\
F_{(3,3,0)} + F_{(3,1,1)} &=& -\frac{175}{768} (3,9) -\frac{75}{256} (5,7)-\frac{39}{128} (7,5)-\frac{223}{768} (9,3)-\frac{83}{256} (11,1) 
\labell{eq:Resumc}
\end{eqnarray}
It is obvious from equations~\reef{eq:Resuma},~\reef{eq:Resumb}
and~\reef{eq:Resumc} that it is easy to predict the terms that will be
present in higher order terms. The coefficients can then be determined
by fitting to the form of the expansion. However, we feel that it
should also be possible to determine the relative coefficients within
any given contribution from some simple underlying set of rules, since
the resummations in \reef{eq:Resuma}--\reef{eq:Resumc} are rather
messy.  By this we mean that, a surface's specific contribution to the
free energy should be able to be isolated, up to an overall constant.
A natural hypothesis arises by considering the diagrams associated
with $\langle b,f \rangle$ as in some way \emph{fundamental} (see
figure \ref{fig:fundamental}); and then trying to build handled
diagrams by \emph{stitching} these fundamental diagrams together by
joining boundaries. The obvious guess for how to represent the joining
of two surfaces, $\langle b_1, f_1 \rangle$ and $\langle b_2, f_2
\rangle$, together is multiplication: $\langle b_1, f_1 \rangle
\langle b_2, f_2 \rangle$.

\begin{figure}[ht]
\begin{center}
\includegraphics[scale=0.22]{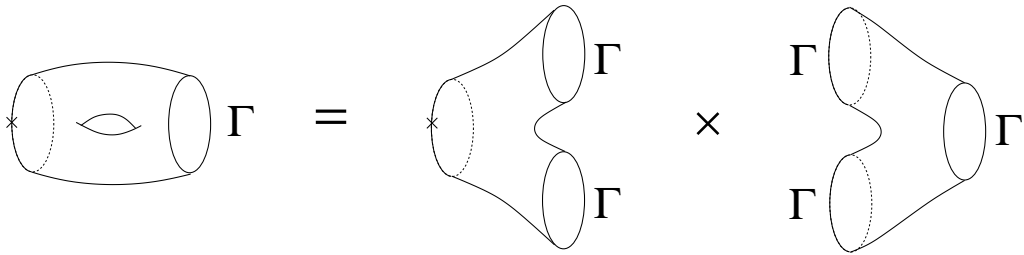}  
\end{center}
\caption{\small Schematic representation of the stitching of diagrams  associated to multiplication to recover terms in the free energy. (See text for details.)} 
\label{fig:multiply}
\end{figure}

It is clear from the form of equations
\reef{eq:Resuma}---\reef{eq:Resumc} that, for this to work, one must
only stitch like boundaries together: that is ZZ boundaries to ZZ
boundaries and FZZT boundaries to FZZT boundaries.  Notice that if $f$
is zero then surfaces with handles have no $\lambda$ dependence; and
if $b$ is zero then surfaces with handles only depend on the
combination $(x + \lambda)$. If these handled terms are formed from
the stitching of fundamental surfaces then those with $f=0$ must be
formed exclusively from ZZ--ZZ stitchings; and those with $b=0$
exclusively from FZZT--FZZT stitchings. Let us explore this by
attempting to construct surfaces with $f=1$. Since $b$ is arbitrary
these surfaces are close in structure to to the $f=0$ case and so we
will try to construct them (at least initially) from ZZ--ZZ stitchings
only. Consider the combined $(1,3,0)$ and $(1,1,1)$ term. Equation
\reef{eq:GenRel} tells us the expected form of the former term, and we
can denote it as $\langle 1,3\rangle$.  Inspection of figure
\ref{fig:multiply} tells us that the $(1,1,1)$ term can only be
constructed from the stitching of two pairs of ZZ boundaries between
the fundamental three--string surfaces $\langle 3,0 \rangle$ and
$\langle 2,1 \rangle$.

Our guess is therefore:
\begin{eqnarray} \labell{eq:111}
F_{(1,3,0)} + F_{(1,1,1)} = a_0 \langle 1,3 \rangle + b_{00} \langle 2,1 \rangle \langle 3,0 \rangle = a_0 (1,3)' + b_{00} (2,1)(3,0)  
\end{eqnarray}
Comparison with equations \reef{eq:Resuma} yields the result
$a_0=5/48, \, b_{00}=-1/32$.  Moving on to the combined $(1,4,0)$ and
$(1,2,1)$ term we predict:
\begin{eqnarray} \labell{eq:211}
F_{(2,3,0)} + F_{(2,1,1)} &=& a_1 \langle 2,3 \rangle + b_{10} \langle 2,1 \rangle \langle 4,0 \rangle + b_{11} \langle 3,0 \rangle \langle 3,1 \rangle \nonumber \\ &=& a_2 (2,3)^{\prime \prime} + b_{10} (2,1)(4,0)^\prime + b_{11} (3,0)(3,1)^\prime  
\end{eqnarray}
The justification for this ansatz is that it contains every pair of
constituent surfaces that could possibly contribute. Again, our
prediction can be realised with $a_1=5/96, \, b_{10}=b_{11}=-1/32$. At
the next level we have:
\begin{eqnarray} \labell{eq:311}
F_{(3,3,0)} + F_{(3,1,1)} &=& a_2 \langle 3,3 \rangle + b_{20} \langle 2,1 \rangle \langle 5,0 \rangle + b_{21} \langle 3,1 \rangle \langle 4,0 \rangle + b_{22} \langle 3,0 \rangle \langle 4,1 \rangle \nonumber \\ &=& a_2 (3,3)^{\prime \prime \prime} + b_{20} (2,1)(5,0)^{\prime \prime} + b_{21} (3,1)^{\prime} (4,0)^\prime + b_{22} (3,0) (4,1)^{\prime \prime} \nonumber \\
\end{eqnarray}
This time we find that $a_2=5/288, \, b_{21}= 2 b_{20}= 2 b_{22} = -1/32$. We can now see a possible pattern emerging amongst the $b_{ij}$:
\begin{eqnarray} \labell{eq:Binomial}
b_{ij} = b_i \left( \begin{array}{c} i \\ j \end{array} \right) \qquad b_0 = -\frac{1}{32} \, , \; b_1 = -\frac{1}{32} \, , \; b_2 = -\frac{1}{64} 
\end{eqnarray} 
Testing this conjecture for the combined $(4,3,0)$ and $(4,1,1)$ term
we find that it does indeed hold, with $a_3=5/1152, \, b_3 = -1/192$.
So it seems that we have uncovered a general rule:
\begin{eqnarray} \labell{eq:GenRule}
F_{(k,3,0)} + F_{(k,1,1)} = a_k \langle k, 3 \rangle + b_k \sum_{i=0}^k \left( \begin{array}{c} k \\ i \end{array} \right) \langle 3+i, 0 \rangle \langle 2+k-i, 1 \rangle    
\end{eqnarray} 
The origin of the binomial factors here is not known. It may relate to
the number of ways that the constituent worldsheets can be stitched
together, but we have not been able to how this would work. Notice
that upon including the $z$ derivatives in \reef{eq:GenRule}, by
replacing the $\langle \, , \, \rangle$ with $( \, , \, )$ everywhere,
the resultant equation is reminiscent of the Liebniz rule for the
multiple differentiation of a product.  Note also that it seems that
we can now associate the $a_i$ as being the coefficients of the
$(i+1,3,0)$ and the $b_i$ as being the coefficients of the
$(i+1,1,1)$. Matters are not this simple however, as will be shown
when we analyze other surfaces below.

We return now to the combined $(1,4,0)$ and $(1,2,1)$ term. Since the
handled term here fits into the $b=1$ subclass analogous to the $f=1$
subclass explored above, we make the natural guess that it can be
constructed from exclusively FZZT-FZZT stitchings. It turns out that
this works providing that we add in an extra term proportional to
$(3,2,1)$. This can be thought of as self-stitching of two ZZ
boundaries on the worldsheet to form a handle. We find:
\begin{eqnarray}
F_{(1,4,0)} + F_{(1,2,1)} &=& A_1 \langle 1,4 \rangle + B_{10} \langle 1,2 \rangle \langle 0,4 \rangle + B_{11} \langle 0,3 \rangle \langle 1,3 \rangle + C_1 \langle 3,2 \rangle  \nonumber \\ &=& A_1 (1,4)^{\prime \prime} + B_{10} (1,2)(0,4)^\prime + B_{11} (0,3)(1,3)^\prime + C_1 (3,2)^{\prime \prime} \ , 
 \labell{eq:121}
\end{eqnarray}
where $A_1=1/48 , \, B_{10}= B_{11}= -1/32 , \, C_1=1/96$. This
seems to be the only sensible choice: all other schemes we tried
yielded coefficients as unilluminating as those in the original
resummation equations \reef{eq:Resuma},~\reef{eq:Resumb}
and~\reef{eq:Resumc}. We can make some sense of this result if
consider the action of the transformation $\langle b, f \rangle
\leftrightarrow \langle f, b \rangle$ on the resummed contributions.
We will denote the action of this transformation on a surface
$(b,f,h)$ as $(b,f,h)^{\ddagger}$. Note that this transformation is
defined at the level of the $\langle b, f \rangle$ themselves, so the
relation $(b,f,h)^{\ddagger} \equiv (f,b,h)$ is not trivially
satisfied! Indeed, we can immediately see a counter example using
\reef{eq:GenRel2}: $F_{(0,0,h)} \propto (6h-6,0) \neq
F_{(0,0,h)^{\ddagger}} \propto (0, 6h-6)$. These $(0,0,h)$ terms are
from the background $u(z)$ expansion of course: perhaps terms from the
$v(z)$ expansion do satisfy $(b,f,h)^{\ddagger} \equiv (f,b,h)$? From
\reef{eq:GenRel} we clearly see that surfaces without handles
certainly do. By requiring that $(1,2,1)^{\ddagger} = (2,1,1)$ we can
unambiguously separate the handled terms from the non-handled terms
(using \reef{eq:121} and \reef{eq:211}):
\begin{eqnarray}
F_{(2,1,1)} &=& \frac{1}{96} \langle 2,3 \rangle - \frac{1}{32} \langle 2,1 \rangle \langle 4,0 \rangle - \frac{1}{32} \langle 3,0 \rangle \langle 3,1 \rangle \nonumber \\
F_{(1,2,1)} = F_{(2,1,1)^{\ddagger}} &=& \frac{1}{96} \langle 3,2 \rangle - \frac{1}{32} \langle 1,2 \rangle \langle 0,4 \rangle - \frac{1}{32} \langle 0,3 \rangle \langle 1,3 \rangle
\end{eqnarray}
and, as such:
\begin{eqnarray}
F_{(2,3,0)} = \frac{1}{24} \langle 2,3 \rangle \, , \qquad F_{(1,4,0)} = \frac{1}{48} \langle 1,4 \rangle 
\end{eqnarray}
This would explain the appearance of the self-stitched $(3,2,1)$ term
in \reef{eq:121}. Using the same procedure we can render the
components of \reef{eq:111} in manifestly invariant form with
$(1,1,1)^{\ddagger} = (1,1,1)$:
\begin{eqnarray}
F_{(1,1,1)} = \frac{1}{48} \langle 1,3 \rangle &+& \frac{1}{96} \langle 1,2 \rangle \langle 2,1 \rangle +\frac{1}{48} \langle 3,1 \rangle \nonumber \\
F_{(1,3,0)} &=& \frac{1}{12} \langle 1,3 \rangle
\end{eqnarray}
The $(1,1,1)$ term here now represents two self-stitched surfaces;
plus one of each of FZZT-FZZT and ZZ-ZZ stitchings between the
fundamental surfaces $\langle 1,2 \rangle$ and $\langle 2,1 \rangle$.

Whether other surfaces are also invariant under the $b \leftrightarrow
f$ operation remains to be seen. At higher orders things get more
difficult because of mixing of many different terms with various
numbers of handles. The pattern cannot be binomial in every case
because for many surfaces there are more possible fundamental
stitchings that can contribute to the free energy than there are
binomial coefficients. So far we have not been able to complete the
elegant picture presented above for arbitrary $(b,f,h)$, though it is
likely that the final solution will be extremely simple and
symmetrical. This is work in progress. However, if all the surfaces
making up $v(z)$ are indeed invariant under $b \leftrightarrow f$,
then we can automatically generate the $(1,f,1)$ series from the
$(b,1,1)$ series \reef{eq:GenRule}. Regrettably, the next term in the
$(1,f,1)$ series, $(1,3,1)$, also mixes with a two--handle term,
$(1,1,2)$; so without knowing the appropriate two--handle rules we
have been unable to test our conjecture.

\subsection{The Other Regime: Large Negative $z$}
Let us now turn to the regime $z\to-\infty$. Recall that in the case
$\lambda=0$ we saw that the physics arranged itself into that of a
probe FZZT D--brane, now in the background of $\Gamma$ units of flux.
Now we wish to use equation~\reef{eq:miuramore} to develop an
expansion for $v(z)$ at non--zero $\lambda$, and examine its
properties. The expansion we obtain is:

\begin{eqnarray}
v(z) &=&\lambda^{1/2}+{\frac {\nu^2 (4\,{\Gamma}^{2}-1)}{8 \, \lambda^{1/2} {
z}^{2}}}- \frac {\nu^3 (4\,{\Gamma}^{2}-1)}{8 \, {\lambda}{z}^{3}}}-{
\frac {\nu^4 (16\,{\Gamma}^{4}-104\,{\Gamma}^{2}+25)}{{
128 \, \lambda}^{3/2}{z}^{4}}\nonumber\\
&+& \frac{\nu^4 (32 \Gamma^4 -80 \Gamma^2 +18)}{32 \, \lambda^{1/2} z^5} + \frac{\nu^5 (16 \Gamma^4 -56 \Gamma^2 +13)}{32 \, \lambda^2 z^5} \nonumber\\
&-& \frac{\nu^5 (2560 \Gamma^4 - 6400 \Gamma^2 + 1440)}{1024 \, \lambda z^6} 
+ \frac{\nu^6 (64 \Gamma^6 -1840 \Gamma^4 + 4768 \Gamma^2 - 1073)}{1024 \, \lambda^{5/2} z^6} + \cdots
\end{eqnarray}

Once again, we have a remarkable mess, and it is arguably even worse
than the one we had for large positive $z$ at non--zero $\lambda$.
Very interesting is the fact that this expansion does not reduce to
the expansion~\reef{eq:formsofv} that we had at $\lambda=0$. In fact,
at $\lambda=0$ if is singular. We shall have to understand the physics
of this.

To construct the free energy of our probe (the logarithm of the
wavefunction) we are instructed by equation~\reef{eq:waveform} (as we
were previously) to integrate once and divide by $\nu$. Having done
so, we find that the first few terms are:
\begin{equation}
F=\lambda^{\frac12}\frac{z}{\nu}+\frac{(4\Gamma^2-1)}{8}\left(-\frac{\nu}{\lambda^{\frac{1}{2}}z}+\frac{\nu^2}{2\lambda z^2}\right)+\cdots\ ,
  \labell{eq:firstfewterms}
\end{equation}
which fits nicely with the asymptotic form of the wavefunction
following from modified Bessel functions, observed in
equation~\reef{eq:asymptoticbessel}.  Now, nowhere do we see a natural
occurrence of the dimensionless coupling $g_s=\nu/z^{3/2}$, nor do we
see its renormalized cousin ${\tilde g}_s=\nu/(z+\lambda)^{3/2}$, so
we appear to have a puzzle.

The resolution is simply that we need not expect either $g_s$ or
${\tilde g}_s$ to appear as the natural stringy expansion parameter in
this situation. Part of the reason is that $\lambda$ and $z$ appear
very differently in this regime as compared to the large positive $z$
regime. There, even when $\lambda$ was zero, at tree level there was a
natural parameter which gave Boltzmann weight to loops of length
$\ell$, and this was $z$, the tree level part of $u(z)$. Introducing
$\lambda$ brings it in to perform a role already performed by $z$ (at
tree level) and so it simply shifts $z$ as we have seen. In this large
negative $z$ regime, we have completely different behaviour.  The
potential $u(z)$ {\it vanishes} at leading order and so at that order
there is nothing weighting the length of loops when $\lambda=0$. 
So when $\lambda$ is non--zero, there is now a weighting parameter at
tree level, and so the natural loop expansion that it controls need
not have anything to do what that of the $\lambda=0$ case.
Furthermore, it must be disconnected from any expansion developed in
the $\lambda=0$ case since the weight $e^{-\lambda\ell}$ allows loops
of infinite length to dominate when $\lambda=0$ which will not yield a
good expansion. An expansion in $\lambda$ should be expected to be
singular there, and this is what we have seen. To handle physics at
$\lambda=0$, we should expect to resum at finite $\lambda$ to the full
$v(z,\lambda)$ and then, setting $\lambda=0$, develop a new expansion
in a different parameter. This is the origin of the second and fourth
expressions in equation~\reef{eq:formsofv}.

Correspondingly, we shall therefore not expect that the natural
expansion parameter is $g_s$ or ${\tilde g}_s$. Instead, an examination
of the expression for the free energy shows that the natural expansion
parameter is:
\begin{equation}
{\hat g}_s=\frac{\nu}{\lambda^{\frac12}z}\ .
  \labell{eq:newcoupling}
\end{equation}
Notice that this is also a dimensionless combination of the important
parameters in the problem, a combination which is not available when
$\lambda$ vanishes. Inspired by the success of our exploration of the
large positive $z$ regime, we write our free energy as:
\begin{eqnarray}
F&=&{\hat g}_s^{-1}-\left(\Gamma^2-\frac{1}{4}\right)\Biggl[\frac{1}{2}{\hat g}_s^1+\frac{1}{4}{\hat g}_s^2+{\hat g}_s^3\left(-\frac{4\Gamma^2-25}{96}+\frac{\lambda}{z}\frac{4\Gamma^2-9}{16}\right)\nonumber\\
&+&{\hat g}_s^4\left(-\frac{4\Gamma^2-13}{32}+\frac{\lambda}{z}\frac{4\Gamma^2-9}{8}\right) \nonumber\\
&+&{\hat g}_s^5\left(-\frac{(16\Gamma^4-456\Gamma^2+1073)}{1280}-\frac{\lambda}{z}\frac{(4\Gamma^2-9)(4\Gamma^2-61)}{192}+\left(\frac{\lambda}{z}\right)^2\frac{(4\Gamma^2-9)(4\Gamma^2-21)}{32}\right)\nonumber\\
&+&O({\hat g}_s^6)\Biggr]
  \labell{eq:finalfree}
\end{eqnarray}
Rather pleasingly, we see that in this new expansion we again have
worldsheets corresponding to a probe brane in a background quantified by $\Gamma$.

\subsection{General Considerations}
It is curious that the expansion of $v_C(z)$ manages to resum so
nicely, and it is worthwhile to investigate this phenomenon a
little more closely. To do so, we again appeal to \reef{eq:miuramore}.
Let
\begin{eqnarray}
v_C(z)=\sum_{n=0}^{\infty}a_{n}(z)\nu^n , \quad \quad
u_\Gamma(z)=\sum_{n=0}^{\infty}b_{n}(z)\nu^{n}.
\end{eqnarray}
Note that \reef{eq:expansion} immediately shows that for $z\to+\infty$,
\begin{equation}
b_0(z)=z  , \quad b_1(z)=\frac{\Gamma}{{z^{1/2}}}  , \quad
b_2(z)=-\frac{\Gamma^2}{2z^2}  , \quad \ldots ,
\labell{eq:coeffbi}
\end{equation}
and for $z\to-\infty$,
\begin{eqnarray}
 b_0(z)=0, \quad \quad b_1(z)=0, \quad \quad b_2(z)=\frac{4\Gamma^2-1}{4z^2},
\nonumber \\
 b_3(z)=0, \quad
b_4(z)=\frac{1}{8}\frac{(4\Gamma^2-1)(4\Gamma^2-9)}{z^5}, \quad
\ldots .
\end{eqnarray}
It is apparent that for $n>0$, $b_n(z)=b_nz^{-\frac{3}{2}n+1}$
where the $b_n$ are independent of $z$. Note that for the negative
$z$ expansion, $b_{2n+1}=0$. We are thus able to write \reef{eq:miuramore} as
\begin{equation}
\sum_{n=1}^{\infty}b_{n}z^{-\frac{3}{2}n+1}\nu^{n}+b_0(z)+\lambda
=
\bigg(\sum_{n=0}^{\infty}a_{n}(z)\nu^n\bigg)^2-\nu\frac{\partial}{\partial
z} \bigg(\sum_{n=0}^{\infty}a_{n}(z)\nu^n\bigg),
\end{equation}
which must hold for each order of $\nu$. Clearly, this gives
$a_0(z)=\sqrt{b_0(z)+\lambda}$. It is straightforward to show
that, for $n>0$,
\begin{equation}
a_n(z)=\frac{1}{2a_0(z)}\bigg(b_n
z^{-\frac{3}{2}n+1}+\frac{\partial}{\partial z} a_{n-1}(z) -
\sum_{k=1}^{n-1}a_{n-k}(z)a_k(z)\bigg).
\end{equation}
Now, for $z\to+\infty$, $b_0(z)=z$, so $a_0(z)=\sqrt{z+\lambda}$.
It is then clear that $a_n(z)$ will take a specific form,
\begin{equation}
a_n(z)=\sum_{i=0}^{N}c_i z^{r_i}(z+\lambda)^{s_i},
\end{equation}
where the $r_i$ and $s_i$ are integer or half-integer and the
$c_i$ are functions of the $b_i$. To obtain the contribution to
the free energy, we want to divide by $\nu$ and integrate once.
Based on the rules we introduced above, the contributions to the
free energy are manifestly expressions in integer and
half--integer powers of $z$ and $(z+\lambda)$, so we would expect
these integrals to yield expressions that also have this form. In
general, integration will produce expressions with the correct
$z$ and $(z+\lambda)$ dependence, but they will also have
nontrivial logarithmic dependence. On the outset, this is
discouraging because we have developed no rules for producing
logarithmic contributions to the free energy. But upon closer
investigation we notice an interesting pattern: the $b_i$ conspire
in such a way as to exactly cancel the logarthmic dependence in
the integral. For example, in the integral of $a_2(z)$, the
coefficient of the log term turns out to be $b_1^2+2b_2$. But
(\ref{eq:coeffbi}) shows that $b_1^2=-2b_2$, exactly the
relationship needed to cancel this coefficient. We have verified
explicitly that all log dependence cancels up to $n=7$.

We turn next to the $z\to-\infty$ expansion. In this regime, $b_0(z)=\lambda$, so
$a_0(z)=\sqrt{\lambda}$. Furthermore, $b_{2n+1}=0$ and $b_{2n}\neq
0$, which suggests that an expression for $a_n(z)$ will inevitably
depend on the parity of $n$. The first few terms are:
\begin{eqnarray}
a_0(z)=\sqrt{\lambda}, \quad a_1(z)=0, \quad
a_2(z)=\frac{b_2}{2\sqrt{\lambda}z^2}, \quad
a_3(z)=-\frac{b_2}{2\lambda z^3}, \nonumber \\
a_4(z)=\frac{b_2(6-b_2)}{8\lambda^{3/2}z^4}+\frac{b_4}{2\sqrt{\lambda}z^5},
\quad a_5(z)=\frac{b_2(b_2-3)}{2\lambda^2 z^5}-\frac{5b_4}{4
\lambda z^6}.
\end{eqnarray}
It is clear that there is a pattern emerging in the powers of $z$
and $\lambda$. It can be proved by induction that the general form
for $a_n(z)$ is:
\begin{eqnarray}
a_n(z)=\left\{\begin{array}{ll}\sum_{i=0}^{\frac{n}{2}-1} c_i
z^{-n+i}\lambda^{-\frac{n}{2}+\frac{1}{2}+i} \quad &
\mbox{for n even}  \\
\sum_{i=0}^{\frac{n}{2}-\frac{3}{2}} d_i
z^{-n+i}\lambda^{-\frac{n}{2}+\frac{1}{2}+i} \quad &\mbox{for n
odd,}\end{array} \right.
\end{eqnarray}
where the $c_i$ and $d_i$ are constants to be determined by the $b_i$.
Unlike the case in the positive $z$ expansion, the $z$--dependence
here is trivial so we can easily integrate to obtain each term's
contribution to the free energy. Simplifying, we find that
\begin{equation}
\int dz a_n(z) = \Big(\frac{1}{z \sqrt{\lambda}}\Big)^{n-1}
\sum_{i=0}^N C_i \Big(\frac{\lambda}{z}\Big)^i
\end{equation}
where the $C_i$ are again functions of the $b_i$, and
$N=\frac{n}{2}-1$ if $n$ is even and $N=\frac{n}{2}-\frac{3}{2}$
if $n$ is odd. Setting ${\hat g}_s=\nu/(\lambda^{\frac12}z)$ allows us
to write the free energy as:
\begin{equation}
F=\sum_{n=0}^\infty{\tilde g}_s^{\:\; n-1} \sum_{i=0}^N C_i
\Big(\frac{\lambda}{z}\Big)^i
\end{equation}
It is interesting to see a new stringy perturbative regime arise for
non--zero $\lambda$, with a string coupling ${\hat g}_s$ that is
distinct from the previously identified string couplings, ${\tilde
  g}_s$.



\section*{Acknowledgments}
JEC is supported by an EPSRC (UK) studentship at the University of
Durham. He thanks the Department of Physics and Astronomy at the
University of Southern California for hospitality during the course of
this project. JSP thanks the Department for Undergraduate research
support.  CVJ's research is supported by the Department of Energy
under grant number DE-FG03-84ER-40168.

\providecommand{\href}[2]{#2}\begingroup\raggedright\endgroup

\end{document}